\documentclass[10pt, aps, pre, showpacs, showkeys, twocolumn]{revtex4-1}
\usepackage{amsmath}
\usepackage{amssymb}
\usepackage{graphicx}
\usepackage{hyperref}
\usepackage{subfigure}
\usepackage{color}

\begin{document}

\title{Evolutionary Dynamics with Fluctuating Population Sizes and Strong
Mutualism}
\author{Thiparat Chotibut}
\email{Electronic address: thiparatc@gmail.com}
\author{David R. Nelson}
\email{Electronic address: nelson@physics.harvard.edu}
\affiliation{Department of Physics, Harvard University, Cambridge, Massachusetts
02138, USA}
\date{\today}
\begin{abstract}
Game theory ideas provide a useful framework for studying 
evolutionary dynamics  in a well-mixed environment. This approach,
however, typically enforces  a strictly fixed overall population size,
deemphasizing natural growth processes. We study a  competitive Lotka-Volterra
model, with number fluctuations, that accounts for natural  population growth
and encompasses   interaction scenarios typical of evolutionary games. We
show that, in
an appropriate limit, the model describes   standard evolutionary games with
both genetic drift and  overall population size fluctuations. However, there
are also regimes where a varying population size can strongly influence the
evolutionary
dynamics. We focus on the strong mutualism scenario  and demonstrate that
standard evolutionary game theory fails to describe our simulation results.
 We then analytically and
numerically determine fixation probabilities as well as  mean fixation times
using matched asymptotic expansions, taking into account the population size
degree of freedom. These results elucidate
the interplay between population dynamics and evolutionary dynamics in well-mixed
systems.
\end{abstract}
\maketitle

\section{Introduction}
Recent advances in experimental evolution open  new directions for quantitative
studies of   evolutionary dynamics \cite{Elena:2003sf,Desai:2013jp}. In
a well-mixed environment such as  a chemostat or  a shaken test tube,   the
relative
frequency of interacting microbes can be  measured over time. Although
 microbial experiments demonstrate an intricate feedback between 
  evolutionary and population dynamics \cite{Dai:2012wd,Sanchez:2013eu,Griffin:2004qe},
theoretical understanding  is often limited to evolutionary dynamics in
a fixed population size, mostly within the  framework of evolutionary game
theory and population genetics \cite{Nowak:2006uq,Frey:2010fk,Blythe:2007nr,Ewens:2004kx,Gillespie:2010uq,Hartl:1997hb,Lambert:2014aa}.

 In a well-mixed system with infinitely large populations,  evolutionary
game theory prescribes deterministic  time evolution of the  relative frequency
$f_i(t)$ of 
species \textit{i}  by  the   replicator dynamics: 
\begin{equation}
\frac{df_{i}}{dt}=
\left[w_i(\boldsymbol f)-\bar{w}(\boldsymbol f)\right] f_i,
\label{eqn: generalized_replicator}
\end{equation}
where $w_i(\boldsymbol f)$ is the frequency-dependent fitness of species
$i$ , and $\bar{w}(\boldsymbol f)=\sum_{j}f_{j}w_{j}(\boldsymbol
f )$ is the mean fitness of all interacting species \cite{Smith:1982it,Nowak:2006uq,Frey:2010fk}.
The replicator dynamics encapsulate
frequency-dependent natural selection:  a fitter species flourishes  and
a weaker species succumbs to evolutionary forces. 
The fitness of species \textit{i} is often defined as  a constant background
plus the total payoff
from interactions, assumed to be linear in the $\{ f_i(t) \}$,  $w_{i}(\boldsymbol
f) = 1 + \sum_{j} a_{ij}f_{j}$, where
$a_{ij}$ is a phenomenological payoff matrix characterizing  interactions
with species $j.$ For two interacting species, which  is typical  in a competition
experiment \cite{Elena:2003sf,Desai:2013jp} and is the focus of this paper,
the frequency $f(t)$ of  species 1 fully specifies the state of evolutionary
dynamics, since the frequency of species 2 is just $1-f(t)$.  In this case,
the replicator dynamics determines the time evolution
of         $ f(t) $ from Eq. (\ref{eqn: generalized_replicator}) as   
\begin{align}
\begin{split}
\frac{df}{dt} &=[\alpha_1 - (\alpha_1+\alpha_2)f](1-f)f, \\
&\equiv v_{E} (f),
\label{eqn: replicator}
\end{split}
\end{align}
where  $\alpha_1 = a_{12}-a_{22}$ and $\alpha_2 = a_{21}-a_{11}$. 

\begin{figure}[t!]
\includegraphics[width = 1\linewidth]{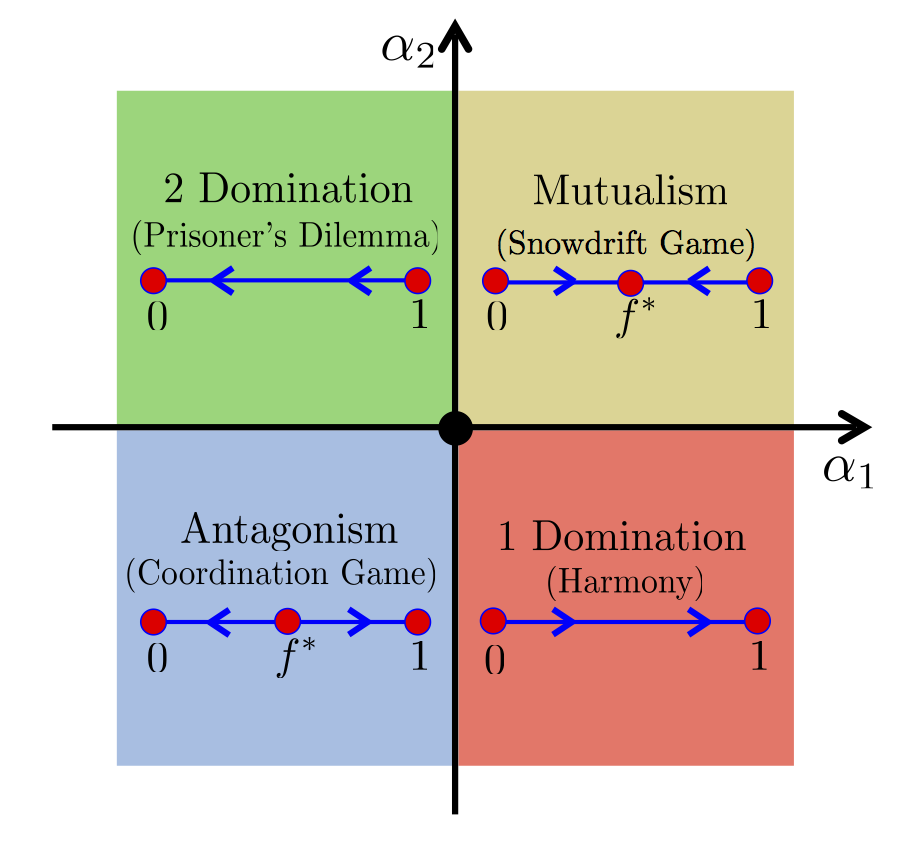}
\caption{\label{fig: Phase_replicator_fixedpop}(Color online) Four competition
scenarios in the  replicator dynamics
represented by the four quadrants.
The fifth scenario, neutral evolution
($\alpha_1 = \alpha_2 = 0$), in which every point $f \in [0,1]$ is a fixed
point, arises at the origin. }
\end{figure}

A rich variety of competition scenarios emerge from this simple description
of evolutionary
games. Depending on the  payoff differences $\alpha_1$ and $\alpha_2,$
Eq. (\ref{eqn: replicator})
exhibits 5 qualitatively different competition scenarios, 
schematically sketched in Fig. \ref{fig: Phase_replicator_fixedpop} \cite{Nowak:2006uq,Frey:2010fk}.
For positive $\alpha$'s (first quadrant of Fig. \ref{fig: Phase_replicator_fixedpop}),
a stable fixed point corresponding to a  species
coexistence appears at $f^* = \alpha_1/(\alpha_1 + \alpha_2),$
lying between the unstable fixed points $f=0$ and $f=1$. This scenario
is commonly referred to
as the ``snowdrift game" in game theory or \textit{mutualism} in the context
of
evolution \cite{Frey:2010fk,Korolev:2011vn,Muller:2014wo}. For
negative $\alpha$'s (third quadrant of Fig. \ref{fig: Phase_replicator_fixedpop}),
the fixed point $f^*$ becomes unstable while the fixed
points with $f$ equals 0 and 1 are stable. This bistability situation is
known by ``coordination
game" in game theory or  \textit{antagonism} in our context. When $\alpha$'s
have opposite
signs (second quadrant and fourth quadrant of Fig. \ref{fig: Phase_replicator_fixedpop}),
scenarios in game theory are either called ``harmony" or ``prisoner's
of the dilemma".
In this case, either $f^*$ or $1-f^*$ exceed unity, and the fixed point
$f^*$ becomes inaccessible. The only 
relevant fixed points are $f=0$ and $f=1$ and only one of them is stable:
For $\alpha_1>0$
 and $\alpha_2<0$, the fixed point $f=1$ is stable and species 1 dominates,
i.e., fixes at 100\% of the population at long times.
For $\alpha_2>0$  and $\alpha_1<0$, the fixed point $f=0$ is stable and
species 2 dominates. Lastly, at the origin of Fig. \ref{fig: Phase_replicator_fixedpop},
when $\alpha_1 = \alpha_2
= 0,$  every point is a fixed
point. We shall refer to this fixed line scenario  as  \textit{neutral evolution},
representing situations when the two interacting species are neutral
variants of each other.

In finite populations, however, evolutionary dynamics are not only influenced
by deterministic frequency-dependent selection term $v_E(f)$, but also by
 randomness
due to discrete microscopic birth and death events, commonly referred to
as number
fluctuations or  genetic drift in population genetics \cite{Hartl:1997hb,Gillespie:2010uq,Ewens:2004kx,Kimura:1984ph}.
Evolutionary game theory in a  strictly
fixed population size can be  reformulated  to account for genetic drift,
with equations that reduce to the deterministic replicator dynamics in the
limit of
infinitely large population size \cite{Nowak:2006uq,Frey:2010fk}. For a
fixed population size $N \gg 1$ and weak payoffs $|a_{ij}| \ll 1$, the
continuum approximation of replicator dynamics with genetic drift reads
\begin{equation}
\frac{df}{dt} = v_E(f)+\sqrt{\frac{2D_g(f)}{N}}\Gamma(t),
\label{eqn: korolev_stoch_replicator}
\end{equation}  
where $\Gamma(t)$ is a zero-mean Gaussian white-noise with a unit variance,
and $D_g(f)/N=f(1-f)/N$ is the frequency-dependent noise amplitude describing
the discrete birth and death processes \cite{Korolev:2011vn}.
In population genetics, the stochastic differential equation (\ref{eqn: korolev_stoch_replicator})
must be interpreted according to Ito's prescription \cite{Korolev:2011vn,Korolev:2010fk},
which we shall assume also for all noise terms appearing in this manuscript.
  For neutral evolution, $v_E(f) = 0$ and the  dynamics is equivalent to
  the continuum limit of the Wright-Fisher sampling or the Moran
process in population genetics, up to a non-universal constant in the noise
amplitude depending
on the  definition
of population size and generation time \cite{Kimura:1962ez,Gillespie:2010uq,Ewens:2004kx,Moran:1962pi,Korolev:2010fk}
that can be absorbed into $N.$
For  $\alpha_1=-\alpha_2 \neq 0$, Eq. (\ref{eqn: korolev_stoch_replicator})
resembles  the generalized Moran process with weak selection \cite{Korolev:2010fk}.
In various contexts, Eq. (\ref{eqn: korolev_stoch_replicator})  and  its
generalizations have received increasing attention as   a simplified model
for
studying the interplay between selection and genetic drift, e.g.  the dilemma
of cooperation \cite{Cremer:2009sh},  rare fluctuations in mutualism
\cite{Mobilia:2010fk,Assaf:2010eb}, the crossover from the mean-field behavior
to fluctuations-dominated behavior in  quantum game theory  \cite{Lassig:2002aa},
as well as competition and cooperation
in spatial range expansions
\cite{Korolev:2011vn,Korolev:2010fk,Lavrentovich:2014oj,Lavrentovich:2013ys}.

Although   replicator dynamics with
genetic drift is a useful approach,  
 the fixed population size condition has
several drawbacks. First, it imposes an artificial growth constraint:
 the
birth of one species necessitates the death of the other even when the
two species are neutral variants. Furthermore, population size fluctuations
away
from a preferred carrying capacity often arise in laboratory experiments,
as well as in natural environments.  For example, understanding how effectively
compressible oceanic
 flows  affect population genetics of marine organisms such as phytoplankton
and cyanobacteria  \cite{Hutchinson:1961bf,Tel:2005kk,Perlekar:2010lk,Pigolotti:2012tx}
requires a time-dependent description of local population size, determined
by a fluid flow structure.
 Incorporating spatially
dependent population sizes into the evolutionary
dynamics of  Eq. (\ref{eqn: korolev_stoch_replicator}) raises important technical
and conceptual challenges \cite{Pigolotti:2013le,Korolev:2013uk}.

In this paper, with the goal of examining the interplay
between number fluctuations, evolutionary and population dynamics in mind,
we study
a two-species competitive
Lotka-Volterra model,
one that couples the replicator dynamics to the dynamics of population
size. Five deterministic competition and cooperation scenarios similar to
 replicator
dynamics
emerge naturally from microscopic birth and competitive death events. Dynamics
in  finite populations
exhibit  selection, genetic drift, and growth of population size, as well
as population
size fluctuations. 

We  first discuss the limit
when    long-time dynamics is governed by weak population  size fluctuations
around
a \textit{fixed}
 stable 
equilibrium population size. If the two competing species reproduce in the
dilute limit at an equal rate, evolutionary and population
dynamics  approximately
decouple near the equilibrium population size. In this case, the effective
evolutionary
dynamics
near the equilibrium population size is  described by replicator dynamics
with genetic
drift.  Despite population size fluctuations,   Moran model results with
and
without selection are recovered.      Pigolotti \textit{et
al. }utilized this limit to extend Eq. (\ref{eqn: korolev_stoch_replicator})
to study population
genetics  in  aquatic environments, where population size also varies in
both
time and space
\cite{Pigolotti:2013le}.

We then study the limit
when evolutionary and population dynamics are coupled and competitions
take place with  systematically varying population sizes as opposed to fluctuations
around
a fixed   equilibrium population size. We focus on the strong mutualism scenario,
where
conventional replicator dynamics with genetic drift fails to predict
the fixation probability, due to  a strong coupling
between
evolutionary and population dynamics. The problem can be restated as a
far from equilibrium escape problem to absorbing boundaries  from an attractive
fixed point in
a two dimensional phase space. The method of matched asymptotic expansions
produces \textit{both }the fixation probability and the mean fixation time
taking
into account the coupled evolutionary
and population dynamics .   

The paper is organized as follow:
 Sec. \ref{sec:model} presents    the mean-field  and  stochastic
description of the competitive Lotka-Volterra model.   The phase portraits
of the  model and of the replicator dynamics
are compared and contrasted. The emphasis is on parameter values such that
an attractive line of
approximately fixed population size dominates the long-time dynamics. This
limit enables
us to identify the mapping between
the  model and the replicator dynamics. In Sec. \ref{sec:Rep_FlucPop} we
discuss the limit when the
replicator dynamics with genetic drift allows  independent population size
fluctuations, and show that standard population genetics results for the
fixation
probability and the mean fixation time  in different selection scenarios
are recovered.
 In Sec.
\ref{sec:strong_mut},
we demonstrate the failure of replicator dynamics with genetic drift to
describe simulations of strong mutualism with a varying population size.
We then construct
the fixation probability and the mean fixation time allowing an \textit{arbitrary}
initial
population size and an initial frequency from the method of matched asymptotic
expansions. We conclude with a summary and  discussions
in Sec. \ref{sec:conclusion}. Details of analytical calculations are presented
in the appendices: Appendix \ref{sec:appndxA} contains derivations of the
coupled stochastic dynamics between the frequency and the population
size. Appendix \ref{sec:MAE} and Appendix
\ref{sec:plateau_QSD} explain the application of matched asymptotic expansions
to achieve the results of Sec. \ref{sec:strong_mut}.
\section{Competitive Lotka-Volterra Model}
\label{sec:model}
The competitive Lotka-Volterra model  accounts for natural population growth
with limited resources;
each individual of the same species $S_i$\ undergoes a logistic birth and
competitive death
process:
\begin{eqnarray}
S_{i }\xrightarrow{\mu_i} S_i+S_i, \label{Birth} 
\\
S_i + S_i \xrightarrow{\lambda_{ii}} S_i
\label{IntraComp},
\end{eqnarray}
where  $\mu_i$ is the reproduction rate of species $i$, and $\lambda_{ii}$
is the rate of intraspecies competition.
The combination of  (\ref{Birth}), which describes an exponential growth
 of population in abundant resources, and  (\ref{IntraComp}), which dominates
when the population size is large, leads to saturation of population size
at the
carrying capacity  $N_{i}^{*}=\mu_i/\lambda_{ii}. $ Experiments show that
a
logistic growth  model accurately captures the growth dynamics
of a single yeast strain in a well-mixed culture \cite{Pearl:1976fk}. 

Interspecies interactions are  modeled by additional competition
\begin{equation}
S_i + S_j \xrightarrow{\lambda_{ij}} S_j, \label{InterComp}
\end{equation} 
where $\lambda_{ij}$ is the rate  at which species $j$ 
wins in the competition for limited resources with species $i$. In general,
$\lambda_{ij} \neq \lambda_{ji}$ for $i \neq j$ although $\lambda_{ij}$ and
$\lambda_{ji}$ must both be nonnegative in this model. The interaction (\ref{InterComp})
encapsulates
situations when  one species suffers from the presence of the others,
for example, by secretions of toxins or competition for the same resources.
 As we will now show, there are 5 generic competition scenarios  analogous
to replicator dynamics. The  population size, however, is not strictly
fixed in this more general model, since the reactions (\ref{Birth}-\ref{InterComp})
do not conserve the
overall population size.

\subsection{Mean field description}
\label{subsec: 2A(meanfield)}
 In a well-mixed environment with an infinitely large population size,
Eqs. (\ref{Birth})-(\ref{InterComp})
can be regarded as chemical reactions and determine the mean field dynamics
of the
number of species $i$, $N_i,$ as
 \begin{eqnarray}
\frac{d N_1}{dt} &= (\mu_1- \lambda_{11}N_1- \lambda_{12}N_2)N_1
,\label{Eqn:dN1dt}\\
\frac{d N_2}{dt} &= (\mu_2 - \lambda_{22}N_2 - \lambda_{21}N_1)N_2
,\label{Eqn:dN2dt}
\end{eqnarray}
where we set the  reaction volume to 1. Without interspecies competition,
each species $i$\ independently grows up
and saturates at the carrying capacity $N_i^*= \mu_i/\lambda_{ii}$. Although
the carrying capacity of the two species can be different in general, we
focus on the case when $N_1^*=N_2^*=N $ for simplicity.
By introducing $c_i=N_{i}/N$, which represents the number of species $i$
relative to its carrying capacity,       Eqs. (\ref{Eqn:dN1dt}) and (\ref{Eqn:dN2dt})
can be non-dimensionalized to read 
\begin{align}
\frac{1}{(1+s_o)}\frac{d c_1}{d \tilde t} &=  c_1\Big(1-c_1-c_2\Big) + \beta_1c_1c_2,\label{Eqn:dc1dt}\\
\frac{d c_2}{d \tilde t} &=  c_2\Big(1-c_1-c_2\Big) + \beta_2c_1c_2, \label{Eqn:dc2dt}
\end{align}
where $\tilde t$ is  the dimensionless time  $\mu_{2} t ,$ 
$s_{o} $ is the reproductive advantage of species 1 near the origin defined
by $1+s_{o} \equiv \ \mu_1/\mu_2,$
and the interspecies competitions are absorbed into    $\beta_1 \equiv 1-
 \Big(\frac{\lambda_{12}}{\lambda_{22}}\Big)\Big(\frac{\mu_2}{\mu_1}\Big)$
 and $\beta_2 \equiv 1-  \Big(\frac{\lambda_{21}}{\lambda_{11}}\Big)\Big(\frac{\mu_1}{\mu_2}\Big).
$    
Note that the $\{ \beta_i \}$ can not exceed unity if $\{ \mu_i \}$ and $\{
\lambda_{ij} \}$ are positive. Three
dimensionless parameters $s_{o}, \beta_1, \beta_2$ control the phase portraits
 in the $c_1$-$c_2$ plane, which always contain \textit{at least}
3  physically relevant fixed points at $(0,0), (1,0)$, and $(0,1)$, corresponding
to the total extinction, the saturation of species 1, and the saturation
of species 2, respectively.
The fixed point $(0,0)$ is always unstable with the  straight heteroclinic
trajectories connecting $(0,0)$ to $(1,0)$ and $(0,0)$ to $(0,1)$   
  describing the logistic growth  of a single species in the absence of
the other.

Two dimensionless parameters $\beta_1$ and $\beta_2$ dictate  competition
scenarios similar to those described by $\alpha_1$ and $\alpha_2$ in the
 replicator
dynamics,
provided an initial condition contains  non-zero population of both species.
However, the overall population size is now allowed to change.
These mean field competition scenarios are illustrated in Figs. \ref{fig:
phase_s1_ct1},   \ref{fig:
phase_s0_ct1}, and \ref{fig: phase_VaryingPop}. 
If the product $\beta_1 \beta_2 < 0,$ the
species \textit{i} with positive $\beta_i$ dominates.
The fixed point corresponding to the saturation of the dominating
 species is stable and the fixed point corresponding to the saturation
of the extinct species is a saddle point.  

When $\beta_1 \beta_2 > 0, $  a fourth dynamically relevant fixed point appears
at  $\boldsymbol c^*
= \frac{1}{\beta_1+
\beta_2 - \beta_1 \beta_2}(\beta_1, \beta_2)$.
If both $\beta_1$ and $\beta_2$ are  negative, we
have a bistable situation similar to \textit{antagonism.} Initial conditions
that lie on the basin of attraction of the fixed point $(1,0)$ and $(0,1)$
result in the total domination (i.e., fixation) of species 1 and species
2, respectively.
The coexistence fixed point $\boldsymbol c^*
$ is a saddle point whose  stable 1-d manifold consists of   the
separatrices such as the trajectory connecting $(0,0)$ to $\boldsymbol c^*$.
Here, coexistence is fragile
and only possible for initial conditions lying exactly on these separatrices.

When both $\beta_1$ and $\beta_2$ are  positive, 
 stable coexistence emerges at the stable  fixed point
$
\boldsymbol c^*
$  similar to \textit{mutualism}. Although we shall refer to this   scenario
as mutualism
to conform to Refs. \cite{Korolev:2011vn} and \cite{Pigolotti:2013le},
we
emphasize that interspecies  interactions actually arise from underlying
competitive interactions. In our case, interspecies  interactions
reduce the growth rate per capita of both species and  restrict $\lambda_{ij}
> 0$ or equivalently  $\beta_{i} <1. $  Stable coexistence
can persist despite the
competition.  The population size
 at $\boldsymbol c^*$, however,  reduces to $\frac{\beta_1 + \beta_2}{\beta_1+
\beta_2 - \beta_1 \beta_2}N$ relative to the upper bound $ 2N$ attained in
the
absence of interspecies competition ($ \lambda_{12}=\lambda_{21}=0,$ or equivalently
$\beta_1 = \beta_2
= 1.$) 

Lastly,  the exceptional case $\beta_1 = \beta_2 = 0$ resembles neutral evolution
such that every point on a one dimensional line $c_{1} + c_{2}= 1$ is a
fixed point. We shall call this fixed line scenario \textit{quasi-neutral
evolution}
as the
two species will not be neutral variants in the dilute limit if $s_o \neq
0$:  A reproductive
advantage near the origin
does not destroy the coexistence line  $c_{1} + c_2 = 1,$  but instead modifies
the
relative abundance of the two species as  population size grows and saturates
somewhere on the fixed line
$c_{1} + c_2 = 1. $ The next subsection discusses the approach toward
population size saturation.

\subsection{Growth of population size and mapping to deterministic replicator
dynamics
when $|\beta_i| \ll 1$}
\label{subsec: 2B(replicator_map)}
\begin{figure*}[ht!]
\includegraphics[width = 1\linewidth]{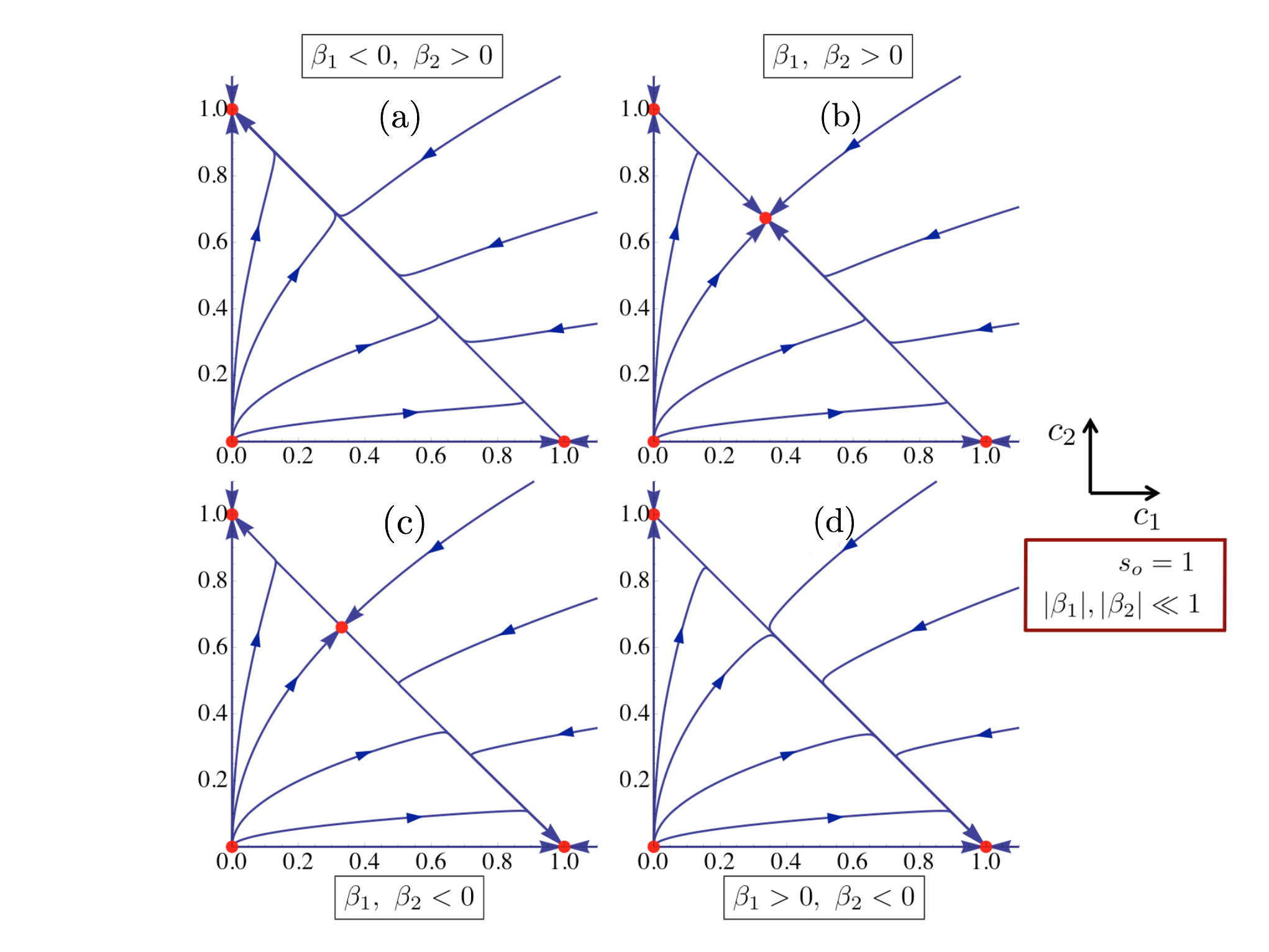}
\caption{\label{fig: phase_s1_ct1}(Color online) The phase portraits of 
competitive Lotka-Volterra
dynamics with
$s_{o} = 1$, $|\beta_2| = 0.028$ and $|\beta_1| = 0.014$ in the $(c_1,c_2)$-plane.
 The competition
scenario
depends on the sign of $\beta_1$ and $\beta_2$,  similar to $\alpha_1$ and
$\alpha_2$ in Fig. \ref{fig: Phase_replicator_fixedpop}. Cases (a),(b),(c),
and (d) correspond to species 2-domination,
mutualism, antagonism, and species 1-domination, respectively. Red circles
represent fixed points and blue lines correspond to  mean field trajectories
of Eqs. (\ref{Eqn:dc1dt})
and (\ref{Eqn:dc2dt}) solved numerically.   For $s_o=1$,
population
size relaxes toward the replicator condition ($c_T \approx 1)$ along a curved
trajectory
of constant $\rho,$ which  forms an upper branch of the parabola  $c_1
= \rho^2 c_2^2 $. Deviation from a trajectory
of fixed $\rho$ only becomes apparent close to the line $c_T=1$. Once
the replicator line $c_1+c_2 \approx 1$ is reached,
the replicator dynamics at a fixed population size takes over. }
\end{figure*}
Despite the rough similarity of the scenarios above to those of replicator
dynamics, the competitive Lotka-Volterra
model contains the overall population size as a dynamical variable. In general,
growth
and competition together do not conserve the population size, as illustrated
in Fig. \ref{fig: phase_VaryingPop}. In the limit
$|\beta_1| \ll 1 $ and  $|\beta_2| \ll 1, $ however, there is an attractive
1-d manifold of approximately fixed population size $c_{1}+c_{2} \approx
1$ on which conventional replicator dynamics determines the ultimate outcome.
We shall refer to the competition near the  line $c_1 + c_2 = 1$ in this
limit as  the competition under the \textit{replicator condition.} Under
the replicator condition, the balance between growth and competitive death

results in an effective replicator dynamics with an approximately fixed 
population size, which we discuss below. Figs. \ref{fig: phase_s1_ct1}
and \ref{fig: phase_s0_ct1} illustrate  competitions  in this limit.
 
We now discuss the growth of population size toward the replicator condition
 and the eventual mapping onto the replicator dynamics.  Upon using $c_{T}
\equiv c_1 + c_2$ to measure the overall population size and defining $f
\equiv c_{1}/c_T$ as the
frequency of species 1, we obtain the following coupled dynamics of $c_T$
and $f$ from Eq. (\ref{Eqn:dN1dt}) and Eq. (\ref{Eqn:dN2dt}),   
\begin{align}
\frac{d c_T}{d\tilde t} &= (1+s_of)v_{G}(c_{T}) +(\alpha_1+
\alpha_2)f(1-f)c^{2}_T,\label{Eqn: MeanField_cT} \\
\frac{df}{d\tilde t} &=  v_E(f) + s_{o} f(1-f)(1-c_T), \label{Eqn: MeanField_f}
\end{align}
where the function $v_{G}(c_T) \equiv c_T(1-c_T)$ in Eq. (\ref{Eqn: MeanField_cT})
describes the logistic
growth of population
size, and the evolutionary dynamics term
in Eq. (\ref{Eqn: MeanField_f}) $v_E(f) \equiv [\alpha_1 + (\alpha_1
+\alpha_2)f]f(1-f)$  resembles the frequency-dependent selection in Eq. (\ref{eqn:
replicator})
 with the identification
\begin{equation}
\alpha_1 = (1+s_o)\beta_1, \ \ \ \textrm{and} \ \ \ \alpha_2
= \beta_2.\label{eqn: map_beta_alpha}
\end{equation} 
We first analyze the quasi-neutral evolution scenario when $\beta_1 = \beta_2
= 0,$ and then treat the case  $0<|\beta_i| \ll 1 $ as a weak perturbation.
In the quasi-neutral scenario, $\alpha_1 + \alpha_2 = 0$ and $c_T$ obeys
 $\frac{d c_T}{d\tilde t} = (1+s_of)v_{G}(c_{T}).$  For any non-zero initial
population size, $c_T$
eventually saturates at  $c_T = 1, $  which
is an attractive 1-d manifold of  fixed points in the original $(c_{1},c_2)$
phase space. As the population size grows from $c_T(0)<1$ or declines from
$c_T(0)>1$ to saturate at $c_T =1$, $c_1(t)$ and $c_2(t)$ change to conserve
the variable $\rho$ defined
by
\begin{align}
\begin{split}
\rho &\equiv c_{2}(t) / c_1(t)^{1/(1+s_o)}, \\
&= c_{2}(0) / c_1(0)^{1/(1+s_{o})},
\label{eqn: Rho_Defn}
\end{split}
\end{align}  
 because Eq. (\ref{Eqn:dc1dt})
and Eq. (\ref{Eqn:dc2dt}) with $\beta_1 = \beta_2 = 0$ implies $d\rho/d\tilde
t=0.$ To  see how the frequency of each species changes as the population
size approaches $c_T = 1$,
it's helpful to rewrite $\rho$ in terms of $f$ and $c_T $ as
\begin{equation}
\rho= c_T(t)^{s_{o}/(1+s_{o})} [1-f(t)]/f(t)^{1/(1+s_{o})}.\label{Eqn: Rho_f}
\end{equation}
Since $\rho$ is a conserved variable,  Eq. (\ref{Eqn: Rho_f}) implies that
the frequency of a reproductively advantageous species increases as $c_T(t)$
grows toward $c_T = 1$ when $c_T(0) \ll 1.$  On the
other hand,  the frequency of a reproductively advantageous species decreases
as $c_T(t)$
declines  toward $c_T = 1$ when $c_T(0) \gg 1$.
If both species grow up at an equal rate ($s_o = 0$), the frequency of each
 is independently
conserved, regardless of $c_T(t)$.  
 
For $0<|\beta_1| \ll 1$ and $0<|\beta_2| \ll 1$, the dynamics of
population size away from $c_T = 1$ still  obeys  $\frac{d c_T}{d\tilde
t} \approx (1+s_{o}f)v_{G}(c_{T})$ since $(1+s_{o}f)v_{G}(c_{T}) \gg [\alpha_1+
\alpha_2]f(1-f)c^{2}_T$ in Eq. (\ref{Eqn: MeanField_cT}). Moreover, the approach
toward $c_T=1$
again follows a trajectory of approximately constant $\rho$  since,
away from $c_T = 1$,
\begin{align*}
\begin{split} 
\left|\frac{d}{d\tilde
t} \ln \rho\right| &=   \Big|c_T[(\beta_2 - \beta_1) +
(\beta_2+\beta_1)f]\Big| \\
&\ll \Big|(1+s_{o}f)(1-c_T) \\
&\hspace{5mm} +[(1+s_{o})\beta_1+\beta_2]f(1-f)c_T \Big| \\
&=  \left|\frac{d}{d\tilde t}
\ln c_T \right  |.  
\end{split}
\end{align*}  
Once $c_T$ is in close proximity
to  1$, \rho$ is no longer approximately conserved. The thin neighborhood
of $c_T
= 1 $ in which conservation is strongly violated, however, becomes vanishingly
small
in the limit $|\beta_i| \ll 1$. Accordingly, we can set $c_T = 1$ in Eq.
(\ref{Eqn: MeanField_cT}) and Eq. (\ref{Eqn:
MeanField_f}) to find in this neighborhood
\begin{equation}
\frac{dc_T}{d\tilde t} \approx 0 \ \ \ \textrm{and}  \ \ \ \ \frac{df}{d\tilde
t} =v_E(f),
\label{eqn: replicator_fct}
\end{equation}
which reproduces deterministic replicator dynamics of a fixed population
size $N$.
The mean
field trajectories in Figs. \ref{fig: phase_s1_ct1} and \ref{fig: phase_s0_ct1}
depict the approach toward the replicator condition in  which
the replicator dynamics at $c_T = 1$ determines how the frequency
of each species changes. Fig. \ref{fig: phase_s1_ct1} illustrates  growth
along the bent trajectories $c_1(t) = \rho^2 c_2(t)^2$ that arises when species
1 has a reproductive
advantage near the origin $(s_o=1), $ while Fig. \ref{fig: phase_s0_ct1}
depicts
growth
along a set of straight lines of fixed species' frequency when
$s_o =0$.

\subsection{Stochastic dynamics}
\label{subsec: (2C)stochastic}

\begin{figure*}[th!]
\includegraphics[width =\textwidth]{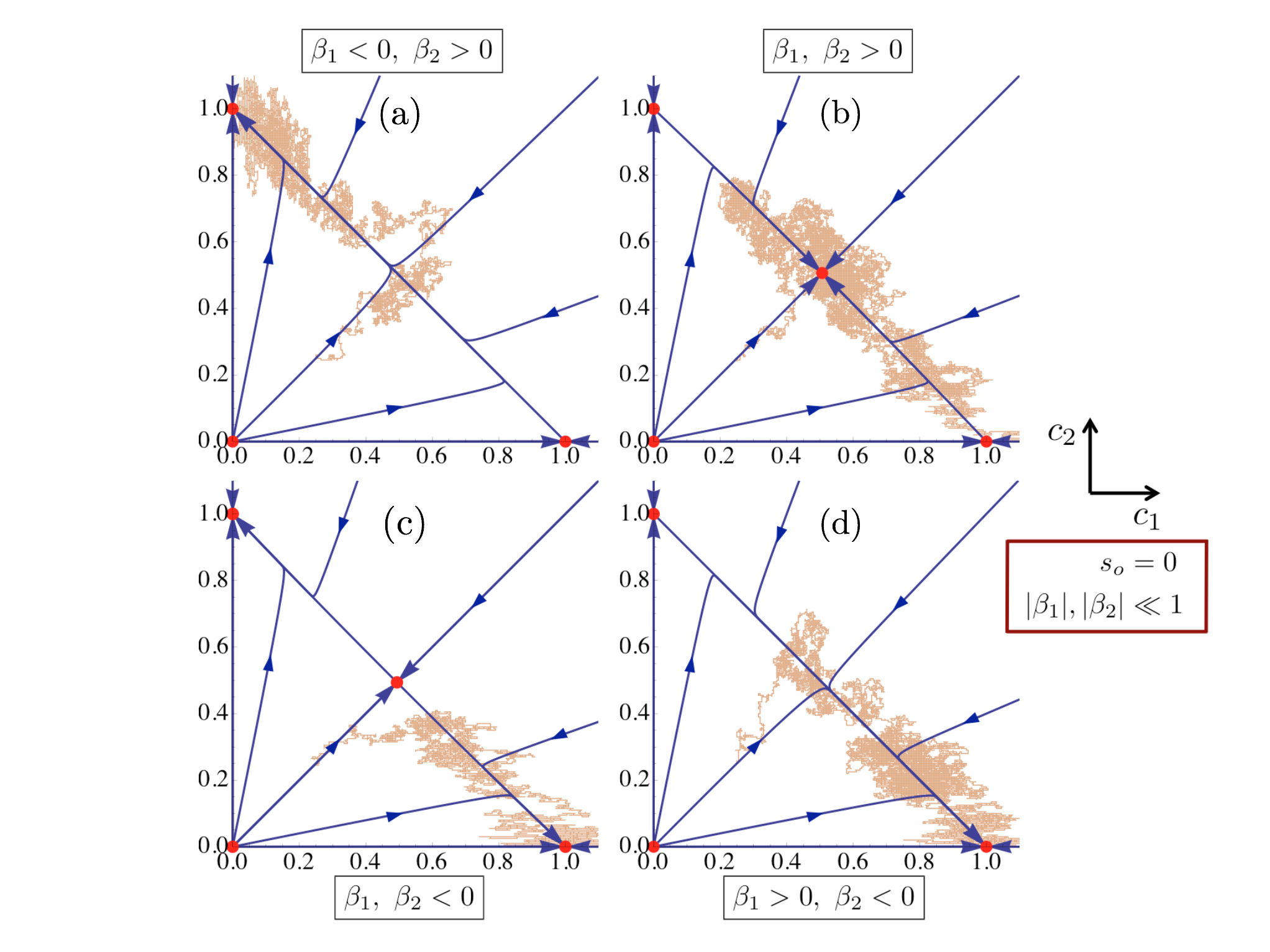}
\caption{\label{fig: phase_s0_ct1}(Color online) Replicator dynamics with
genetic drift and population size fluctuations when $c_T \approx 1$. Cases
(a),(b),(c),
and (d) again correspond to species 2-domination,
mutualism, antagonism, and species 1-domination, respectively with
$s_{o} = 0$, and $|\beta_{1}|=|\beta_2| =0.03.$ In this case, the  mean
field
trajectories approach the replicator condition ($c_1 + c_2 = c_T \approx
1$) as straight lines  that preserve
the initial species frequency
$f$. Each orange  stochastic trajectory, simulated from the Gillespie's
algorithm with the initial condition
$(f,c_{T})=(0.5,0.5)$ (i.e., $c_1 = c_2 = 0.25$)  and $N = 100 $ individuals,
demonstrates   a typical fixation event.  Stochastic trajectories show rapid
growth of population size toward the
replicator condition where selection,  genetic drift, and population size
fluctuations ultimately determine the competition outcome. }
\end{figure*}

\begin{figure*}[th!]
\includegraphics[width = \textwidth]{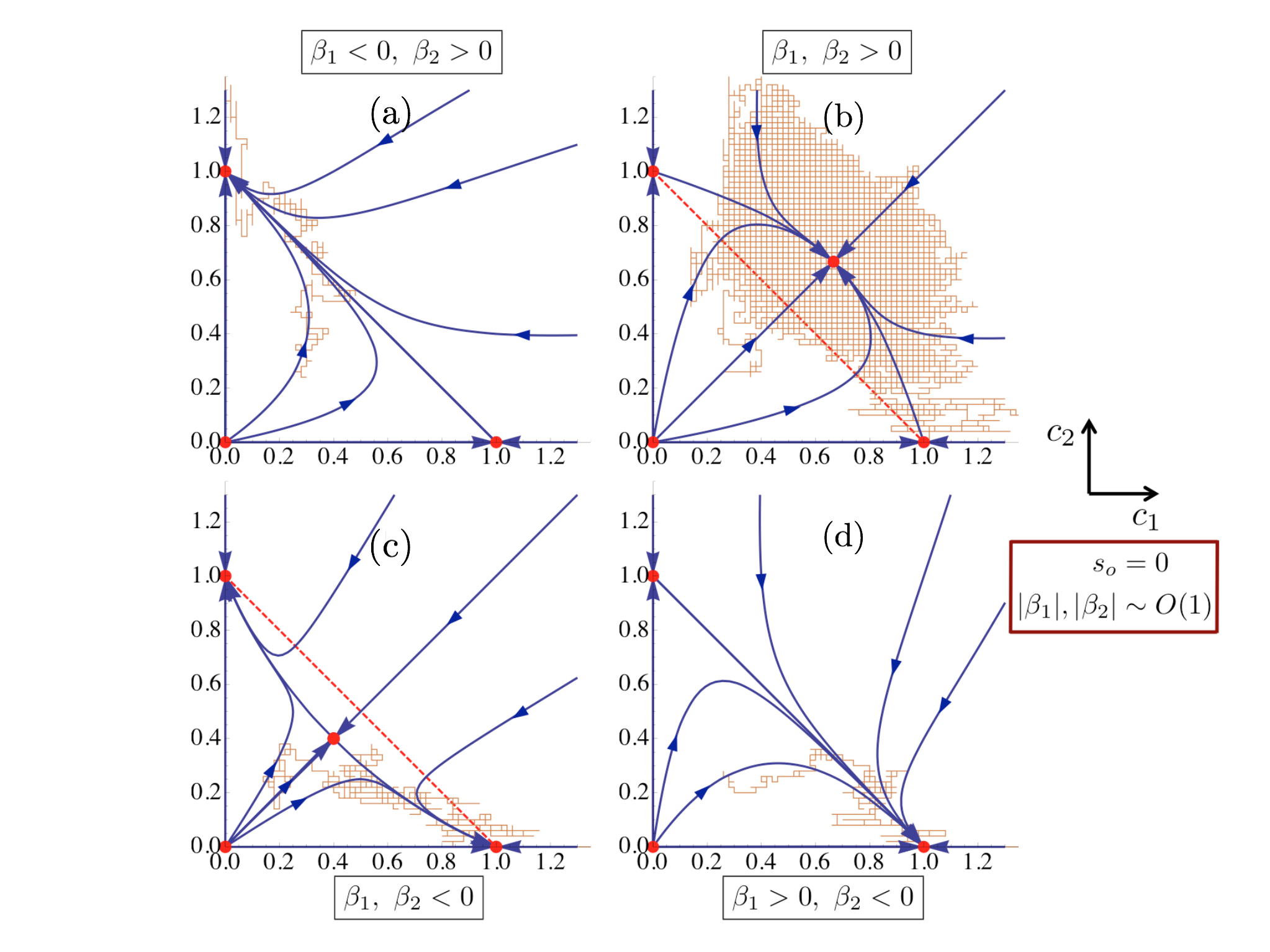}
\caption{\label{fig: phase_VaryingPop}(Color online) The phase portraits
for  
$s_{o} = 0$, and $|\beta_{1}|=|\beta_2| =0.5, $  which includes the case
of strong mutualism of Sec.
\ref{sec:strong_mut}. Competition at long times
no longer takes
place close to the line $c_1 + c_2 = c_T = 1$, depicted as the red dashed
line in (b) and (c), but with a varying overall population size. Cases (a),(b),(c),
and (d) correspond to species 2-domination,
mutualism, antagonism, and species 1-domination, respectively. The initial
condition
for the stochastic simulation is
$(f,c_{T})=(0.5,0.5)$ (i.e., $c_1 = c_2 = 0.25$)  with $N = 50$ individuals.
In contrast to mutualism under the
replicator condition (e.g., Fig. \ref{fig: phase_s0_ct1}(b) with $| \beta_1|,
| \beta_2 | \ll 1$), the coexistence
fixed point in strong mutualism shown
in (b) is highly stable and fixation  becomes a rare event even when $N$
is as small as 50.  }
\end{figure*}

In finite populations, the ultimate fate of the coupled system
depends not only on the 3 dimensionless parameters $s_{o}, \beta_1, \beta_2$
and the initial condition,
but also on  fluctuation corrections to  the mean-field dynamics due to microscopic
stochasticity. We can quantify the stochastic dynamics by regarding the
microscopic rates in Eqs. (\ref{Birth})-(\ref{InterComp}) as Markov processes.
The joint probability distribution of finding $N_i$ individuals of species
$i$ at time $t$,
$P(N_1,N_2,t)$, then obeys the Master equation  
\begin{align*}
\partial_t P(N_1,N_2,t)= &\mu_1(N_1 - 1)P(N_1-1,N_2,t)\\
&+ \mu_2(N_2-1)P(N_1,N_2-1,t)\\
&+\lambda_{11}N_1(N_1+1) P(N_1+1,N_2,t)\\
&+\lambda_{22}N_2(N_2+1)P(N_1,N_2+1,t)\\
&+\lambda_{12}N_2(N_1+1) P(N_1+1,N_2,t)\\
&+\lambda_{21}N_1(N_2+1)P(N_1,N_2+1,t)\\
&-\Big[\hspace{1mm}\mu_1N_1 +\mu_2N_2\\
&\hspace{2mm}+\lambda_{11}N_1(N_1-1)+\lambda_{22}N_2(N_2-1)\\
&\hspace{2mm}+\lambda_{12}N_1N_2 +\lambda_{21}N_1N_2\hspace{1mm}\Big]P(N_1,N_2,t).
\label{eqn: master_eqn}
\end{align*}
In the limit $1/N \ll 1$ (recall that $N =\mu_1/\lambda_{11} = \mu_2/\lambda_{22}$),
this discrete Master equation can be approximated
by the Fokker-Planck equation for the continuous probability distribution
$P(c_1,c_2,t)    $ via the Kramers-Moyal expansions or the Van-Kampen $1/N$
expansions
\cite{Van-Kampen:1992tk,Gardiner:1985qr}. The corresponding  Fokker-Planck
equation for the probability 
$P(\boldsymbol{c},t)$ of a particular species configuration $\boldsymbol{c}$
reads
\begin{eqnarray}
\partial_t P(\boldsymbol{c},t) = \sum_{i=1}^2 \Big( &&-\partial_{c_i}[v_i(\boldsymbol{c})P(\boldsymbol{c},t)]\nonumber\\
&&+ \frac{1}{2N}\partial_{c_i}^2[D_i(\boldsymbol{c})P(\boldsymbol{c},t)]\Big),
\label{eqn: FP}
\end{eqnarray}  
where the deterministic drift and $N$-independent diffusion coefficients
are
\begin{eqnarray}
v_1(\boldsymbol{c})=\mu_1
c_1(1-c_1-c_2) +\mu_1
 \beta_1 c_1 c_2,\label{eqn: drift1}
\\
v_2(\boldsymbol{c})=\mu_2
c_2(1-c_1-c_2) +\mu_2 \beta_2 c_1 c_2,\label{eqn: drift2}
\\
D_1(\boldsymbol{c})= 
\mu_1 c_1(1+c_1+c_2) -\mu_1
 \beta_1 c_1 c_2 ,\label{eqn: diff1}
\\
D_2(\boldsymbol{c})= 
\mu_2 c_2(1+c_1+c_2) -\mu_2 \beta_2 c_1 c_2 .\label{eqn: diff2}
\end{eqnarray}     
An equivalent representation in terms of the Ito calculus \cite{Gardiner:1985qr}
prescribes stochastic dynamics of the $c_i(t)$ that resembles a set of coupled
Langevin equations:
\begin{equation}
\frac{dc_i}{d t} = v_i(\boldsymbol{c}) +\sqrt{\frac{D_i(\boldsymbol{c})}{N}}\Gamma_i(t),
\label{eqn: langevin_ci}
\end{equation}
where $\Gamma_i(t)$ is a  Gaussian white-noise with $\langle\Gamma_i(t)\Gamma_j(t')
\rangle = \delta_{i,j}\delta(t-t')$ and $\langle \Gamma_i(t) \rangle =
0.$
In the limit of infinitely large population size $N$, the noise term of
order $\sqrt{1/N}$ in Eq. (\ref{eqn: langevin_ci}) vanishes and we recover
the mean-field description of Eq. (\ref{Eqn:dc1dt}) and Eq. (\ref{Eqn:dc2dt}).
Note that the deterministic
drift can not be written as a gradient of a potential function since $|\nabla
\times
\vec{v}(\boldsymbol c)|  \equiv |\partial_1v_2(\boldsymbol c)-\partial_2
v_1(\boldsymbol c)|= |\mu_1(1-\beta_1)c_1-\mu_2(1-\beta_2)c_2|\neq 0.$ In
contrast to  diffusion in a potential
field, the non-potential drift  is typical for stochastic non-linear dynamics
in a higher dimensional phase space  \cite{Dykman:1994pz,Gardiner:1985qr}.
Hence, standard
tools for analyzing the statistics of fluctuations such as eigenfunction
expansions of the Fokker-Planck equation \cite{Risken:1984df} or saddle-point
approximations of the most probable escape path \cite{Langer:1969aa,Van-Kampen:1992tk}
are not directly applicable. 

For finite $N$, number fluctuations  alter
the mean-field description
and can lead to outcomes different from the deterministic predictions. For
instance,  fluctuations
will eventually drive
one of the two species to fixation and destroy  stable coexistence for mutualism.
Regardless of the deterministic
phase portraits, the eventual fate of the system at long times is fixation
of a single
species.  Once one species becomes fixed, the dynamics of the fixed species
 follow  stochastic logistic growth while the other species remains forever
extinct, as is easily checked from Eqs. (\ref{eqn:
drift1})-(\ref{eqn: langevin_ci}). The  $c_1$ and  $c_2$ axes 
  are thus \textit{absorbing boundaries} leading to the fixation
of species 1 and of species 2, respectively. The only  absorbing state
in the phase space is total extinction   (0,0), which is inaccessible since
we
don't allow for the death process $S_i \rightarrow \phi $ in our simulations.
Had we included
the death process, the total extinction would nevertheless be extremely unlikely
because
the mean time  to extinction for  logistic growth grows exponentially
 with $N$ \cite{Ovaskainen:2010bc}.              

If  time is non-dimensionalized to $\tilde t= \mu_2 t$ in Eqs. (\ref{eqn:
drift1})-(\ref{eqn: langevin_ci}),  the
stochastic
dynamics  then depends on only 4 dimensionless parameters:   
$s_o, \beta_1, \beta_2, $ and $1/N$. The parameters   $s_{o}, \beta_1, \beta_2$
 control both the mean-field phase portrait and the  diffusion coefficients,
while the parameter $1/N$  sets the  strength of fluctuations
relative to deterministic drift.

In the following sections, we investigate the dynamics in different ranges
of     $s_{o}, \beta_1, \beta_2$
assuming small fluctuations, $1/N \ll 1$.  In Sec. \ref{sec:Rep_FlucPop},
we study
the limit
when $|\beta_i| \ll 1$ with typical stochastic
trajectories shown in Fig.  \ref{fig: phase_s0_ct1} for  $s_{o}=0$.   In
Sec. \ref{sec:strong_mut}, we highlight the limitations of fixed population
size replicator dynamics by studying
a strong mutualism
scenario where $\beta_1 $ and $\beta_2$ are $\mathcal{O}(1)$ and we set $s_{o}=0$
for simplicity.
Fig. \ref{fig: phase_VaryingPop}  illustrates the stochastic dynamics in
this
case, where the time dependence of the overall population size plays a crucial
role. In this situation, replicator dynamics is \textit{never} an appropriate
description.


 \section{Replicator Dynamics with Genetic Drift and Population Size Fluctuations}
 \label{sec:Rep_FlucPop}

We now
follow Pigolotti \textit{et
al.} \cite{Pigolotti:2013le} and discuss competitive Lotka-Volterra dynamics
under the replicator condition
($c_T \approx 1$ and $|\beta_i | \ll 1$)
and $1/N \ll 1,$ thus extending  Eq. (\ref{eqn: korolev_stoch_replicator})
to include fluctuations in the overall population size. We recast earlier
results of Ref. \cite{Pigolotti:2013le} in the language of conventional replicator
dynamics with genetic drift, to better illustrate the breakdown of this approach
for the case of strong mutualism discussed in Sec. \ref{sec:strong_mut}.
Our focus is on the dynamics
of\  $f(t)$, the frequency of species 1, and the total population size $c_T(t)$.
 When $s_o=0 $ and
$ |\beta_i|
\ll 1,$  (Appendix A also treats $s_0 \neq 0$ a case not considered in Ref.
 \cite{Pigolotti:2013le}) Appendix \ref{sec:appndxA} shows that the coupled
stochastic dynamics
of $f$ and $c_T$
 for $c_T \approx 1$  read \cite{Pigolotti:2013le}
\begin{align}
\frac{df}{dt} &=\mu v_E(f)c_T+\sqrt{\frac{\mu D_g(f)}{N}\left(\frac{1+c_{T}}{c_T}\right)}\Gamma_f(t),\label{eqn:dfdt_s0}\\
\frac{dc_T}{dt} &= \mu v_{G}(c_T)+\sqrt{\frac{\mu c_T(1+c_T)}{N}}\Gamma_{c_T}(t),\label{eqn:dcTdt_s0}
\end{align}
where $\Gamma_i(t)$ is an uncorrelated Gaussian white noise with zero mean
and unit variance, $D_g(f) =  f(1-f)$   
is  the frequency-dependent  genetic drift
coefficient  \cite{Gillespie:2010uq,Hartl:1997hb}, $v_E(f)$
and $v_G(c_T)$ are the selection function and the logistic growth function
that appear in Eqs. (\ref{Eqn: MeanField_cT}) and (\ref{Eqn: MeanField_f}).
These stochastic differential equations, which arise from
a more general dynamics  with $s_o \neq 0$ discussed in Appendix \ref{sec:appndxA},
must be interpreted in terms of  Ito calculus \cite{Van-Kampen:1992tk,Gardiner:1985qr}
in order to correctly reproduce the Fokker-Planck Equation (\ref{eqn: FP}).
 We have retained the original unit of time
to make the reproduction time scale
explicit and  denoted $\mu = \mu_2$  for brevity. 

 In Eq. (\ref{eqn:dcTdt_s0}), the dynamics of population size is $f$-independent
and exhibits a combination of  fast approximately deterministic relaxation
toward
the equilibrium line $c_T=1$ and  slow fluctuations with variance $1/N$ around
this equilibrium.   On the other hand, the dynamics of $f$ depends on $c_T$.
Nevertheless,
 it is mostly influenced by the mean population size    $\langle c_T \rangle
=
1 $ since the variance  of $c_T$ about $c_T=1$ is $1/N \ll 1.$ Thus, the
 effective dynamics
of  $f,$ accurate to first order in $1/N,$ can  be approximated by simply
replacing
$c_T =1,$  which leads to 
\begin{equation}
\frac{df}{dt} = \mu v_E(f)+\sqrt{\frac{2 \mu }{N}D_{g}(f)}\Gamma_f(t).
\label{eqn: replicator_s0}
\end{equation} 
Eq. (\ref{eqn:dcTdt_s0}) and Eq. (\ref{eqn: replicator_s0}) together describe
the dynamics near the replicator condition when $1/N \ll 1,$ which is precisely
Eq. (\ref{eqn: korolev_stoch_replicator}), with the addition of an independently
fluctuating
population size around the fixed
mean  $c_T =1$. 

Note that the variance per generation time of Eq. (\ref{eqn: replicator_s0})
  given by $f(1-f)/N$ is independent of both the population size fluctuations
away from $c_T = 1$ and the selection mechanism in the vicinity of this line.
In fact, the variance resembles that of the Wright-Fisher or Moran
model \cite{Gillespie:2010uq}. Thus, the effective population
size deduced from the variance of the genetic drift is equivalent to the
mean population size
$N$ despite
fluctuations in the overall population size. 

From the closed form Eq. (\ref{eqn: replicator_s0}) for $f(t),$  we  can
recover known results for the fixation probability $u(f)$ and the
mean fixation time  $\tau(f)$ which are, respectively, the probability
that species
1 become fixed (instead of species 2) and the average time to  lose heterozygosity
provided
species 1 initially has
a frequency  $f$ at $c_T = 1$.
 These quantities obey ordinary differential equations,
 \begin{align}
 v_E(f)\frac{d}{df}u(f) +\frac{D_g(f)}{N}\frac{d^2}{df^2}u(f)&=0, \\
 v_E(f)\frac{d}{df}\tau(f) +\frac{D_g(f)}{N}\frac{d^2}{df^2}\tau(f)&=-\frac{1}{\mu},
\end{align}
subject to the boundary conditions $u(0) = 0, u(1) = 1$ and $\tau(0) =
\tau(1) = 0$ \cite{Van-Kampen:1992tk,Gardiner:1985qr}. 
The differential equations can be integrated directly leading to  closed
form solutions which
read
\begin{equation}
u(f) = \frac{\int_0^{f} \, e^{-N\Psi(x)} \, dx}{\int_0^{1} \, e^{-N\Psi(x)}
\, dx}, \label{eqn: fixprob}
\end{equation}
where $\Psi(x)  \equiv \int_0^x v_E(y)/D_g(y) \, dy,$ and 
\begin{equation}
\tau(f) = I(1)u(f)-I(f), \label{eqn: MFT}
\end{equation}
where $I(f)  \equiv [ \int_0^f dx \, e^{-N\Psi(x)}\int_1^x dy \, e^{N\Psi(y)}/D^{}_g(y)](N/\mu).$
 
We now review the implications of Eq. (\ref{eqn: replicator_s0})  for different
selection scenarios. In neutral evolution $(\beta_1 = \beta_2 = 0)$, 
Eq. (\ref{eqn: replicator_s0}) becomes ( with $D_g(f) =  f(1-f)$ )
\begin{equation}
\frac{df}{dt} = \sqrt{\frac{2 \mu }{N}D_{g}(f)}\Gamma_f(t),\label{eqn:dfdt_neutral}
\end{equation}
which is a continuous approximation of the Moran model or the Wright-Fisher
sampling in population genetics \cite{Gillespie:2010uq,Crow:1970jh,Ewens:2004kx,Hartl:1997hb,Kimura:1969pt}.
Only genetic drift participates in the dynamics and fixation events are
results of an unbiased random walk with $c_{T} \approx 1$ toward $f=0$ or
$f=1$,
absorbing boundaries, and independent
fluctuations of population size about the mean $N$.  In this case, direct
evaluation of Eqs. (\ref{eqn: fixprob}) and (\ref{eqn: MFT}) gives
\cite{Crow:1970jh}
\begin{equation}
u_{neutral}(f)=f,  \label{eqn:fixprob_neutral}
\end{equation}
\begin{equation}
\tau_{neutral}(f) =-\left(\frac{N}{\mu}\right)\Big[f\ln f+(1-f) \ln (1-f)\Big],\label{eqn:MFT_neutral}
\end{equation}
where $f$ is the initial frequency of species 1.

For selection that favors domination of one species, the special
case  $ \tilde s \equiv\beta_1 = -\beta_{2}$ reproduces the Moran process
with an effective
selective
advantage $\tilde s$ (provided $c_T \approx 1 $), described by
\begin{equation}
\frac{df}{dt} = \mu \tilde s f(1-f)+ \sqrt{\frac{2 \mu }{N}D_{g}(f)}\Gamma_f(t).
\end{equation}
 We emphasize
that
the growth rates of the two species when $c_T \ll 1$ in this particular competitive
Lotka-Volterra dynamics are strictly  identical ($s_o = 0$), but  the species
with positive $\beta_{i}$ nevertheless behaves
near $c_T = 1$ as if it has a selective advantage $\tilde s$ . Upon evaluating
 Eq. (\ref{eqn: fixprob}), we arrive at
the celebrated
Kimura result for the fixation probability \cite{Kimura:1962ez} 
\begin{equation}
u(f) = \frac{1-e^{-\tilde s N f}}{1-e^{-\tilde s N}}.
\end{equation}A lengthy closed-form formula for the mean fixation time can
also be obtained;
see for example Ref. \cite{Cremer:2009sh}. 

For antagonistic or mutualistic interactions, the effective dynamics of
$f$ reads
\begin{equation}
\frac{df}{dt} = \mu \tilde \beta f(1-f)(f^*-f)+ \sqrt{\frac{2 \mu }{N}D_{g}(f)}\Gamma_f(t),
\end{equation}
where   $f^* = \beta_1/(\beta_1 + \beta_2)$ is the coexistence fixed point
with $c_T \approx 1$  and $\tilde \beta \equiv (\beta_1 + \beta_2) $ controls
the stability of $f^*$. The parameter $\tilde \beta$ is positive and negative
for mutualism
and antagonism, respectively. In either case, the fixation probability
 directly follows
from   Eq. (\ref{eqn: fixprob}), and is given by 
\begin{equation}
u(f)=\frac{\int_0^f e^{\frac{N\tilde \beta}{2}(f^*-f)^2} df}{\int_0^1 e^{\frac{N\tilde
\beta}{2}(f^*-f)^2}df},\label{eqn: FP_MutualismCrowded}
\end{equation} 
in agreement with Ref. \cite{Korolev:2011vn}.
It appears that the  mean fixation time from Eq. (\ref{eqn: MFT}) can not
be simplified further,
and must be evaluated numerically.

Pigolotti \textit{et al.} simulated the fixation probability for different
competition scenarios under the replicator condition with $s_o = 0$ and found
good agreement with  these predictions of the fixation probabilities even
for fairly small  population
sizes of $O(N) \sim 100$ individuals \cite{Pigolotti:2013le}.    Constable
\textit{et al.} also studied  this limit  using a different mathematical
technique and  found good agreement between theories and simulations  of
both the fixation probability and the mean fixation time \cite{Constable:2015aa}.
These results confirm that the competitive Lotka-Volterra model reduces to
  replicator dynamics with genetic drift and an independently fluctuating
overall population
size, \textit{provided} $s_o=0$, $|\beta_i|\ll1$, and $c_T \approx 1$.

We mention briefly that when $s_0 \neq 0 ,$ the long-time dynamics still
fluctuates around the equilibrium line $c_T =1 $ provided $|\beta_i|\ll1;$
however,  evolutionary dynamics now couples to  population  dynamics, see
Appendix \ref{sec:appndxA}.  An interesting  phenomenon of fluctuation-induced
selection arises as a result of this coupling near the equilibrium line.
In
the scenario of quasi-neutral evolution ($\beta_1=\beta_2 =0$),
species with a \textit{reproductive disadvantage}   in the dilute limit acquires
a \textit{selective
advantage }for competitions near the equilibrium line \cite{Parsons:2007aa,Parsons:2008aa,Kogan:2014aa,Lin:2012wf}.
The resulting effective evolutionary dynamics near the equilibrium line contains
not only
a fluctuation-induced selective advantage,
but also an unusual genetic
drift of a non-Wright-Fisher (and non-Moran) type \cite{Parsons:2007aa,Parsons:2008aa,Kogan:2014aa,Lin:2012wf}.


\section{Strong Mutualism with  a Varying Population Size}
\label{sec:strong_mut}
In this section, we study a strong mutualism scenario ($\beta_i \sim \mathcal{O}(1)$
in Fig.  \ref{fig: phase_VaryingPop}(b)),
where
the replicator condition is no longer satisfied. In this limit, the coexistence
fixed
point  shifts far away from the line $c_T =1$ and becomes
strongly attractive in all eigen-directions. The faint orange grid in Fig.
 \ref{fig: phase_VaryingPop}(b)
illustrates
a typical fixation trajectory   exhibiting a decline of overall population
size
as  weak fluctuations  about the strongly stable fixed point eventually drive
one of the two species (in this case, species 1) to fixation.        

\subsection{Failure of the fixed population size model near boundary layers}
\label{subsec: 5A(mutual_fixprob)}
   Suppose we accept Eq. (\ref{eqn: korolev_stoch_replicator}) as a phenomenological
model for mutualism
and fit the resulting fixation probability in Eq. (\ref{eqn: FP_MutualismCrowded})
to simulation data;
how well would this model with a strictly fixed population size do? To 
motivate the choice of fitting parameters, we first discuss the behavior
of the fixation probability $u(f) $ predicted by Eq. (\ref{eqn: FP_MutualismCrowded}).
For  $\tilde \beta N \ll
1 $ (recall that $\tilde \beta \equiv \beta_1 + \beta_2$ in Eq. (\ref{eqn:
FP_MutualismCrowded})), genetic drift dominates  mutualistic selection and
the fixation probability
approaches the
result of an unbiased random walk of neutral evolution, Eq. (\ref{eqn:fixprob_neutral}).
For     $\tilde \beta N \gg 1 $,
the coexistence fixed point is metastable and fixation driven by weak genetic
drift
becomes a rare   event. Initial conditions in close proximity to $f^*$
almost
surely visit $f^*$ before fixation occurs, giving rise to a plateau
of equal     fixation probability $u(f^*)$ in the neighborhood of $f^*
$.  Furthermore, the fixation probability  $u(f) $ 
only varies rapidly within the boundary layers of width  $\sim1/N$ adjacent
to each of the absorbing states $f=0$ and $f=1$, away from which   $u(f)
$ exhibits
crossovers
to a plateau value  $u(f^*).$  The boundary layers near the absorbing states
contain initial conditions
that can be driven by genetic drift to fixation  before being attracted toward
$f^*$. For
symmetric mutualism ($f^*=1/2$), the plateau height $u(f^*) $ is 1/2 by
 symmetry from Eq. (\ref{eqn: FP_MutualismCrowded}) and is independent of
$N$. For asymmetric mutualism,  the $N$-dependent behavior of $ $$u(f^*)$
  can be understood by studying rare event escape from a metastable
state.  For  an evolutionary game with a stable coexistence fixed point,
  it can be shown that $u(f^*)$ is given by the ratio of the flux into
the absorbing state $f=1$ to the total flux into the absorbing states
$f=0$ and $f=1$ whose  $N$-dependent behavior in the limit $N \gg 1$ is
given by \cite{Assaf:2010eb,Mobilia:2010fk}
\begin{equation}
u(f^*) \approx \frac{1}{1+e^{-N\Delta S_0 + \Delta S_1}},\label{eqn:plateau_height}
\end{equation} 
where, from the perspective of a Feynman's path integral formulation of 
stochastic dynamics \cite{Kleinert:2009aa,Altland:2010aa},  $\Delta S_0 \equiv
S_{0}[\gamma_{f^* \rightarrow 1}]-S_0[\gamma_{f^* \rightarrow 0}]$ is the
difference between the ``action" $S_0[\gamma_{f^{*} \rightarrow x}]$ associated
with the  most probable escape path   $\gamma_{f^{*} \rightarrow
x}$ beginning at $f^*$ and ending at an absorbing state $x$, and $\Delta
S_1\equiv \ln w[\gamma_{f^{*} \rightarrow
1} ]-\ln w[\gamma_{f^{*} \rightarrow
0} ] $\ 
is the difference between  fluctuations corrections$ $ to the  action
of  the most probable escape path.  The  $N$-independent
functions   $\Delta
S_0$ and $\Delta S_1$ are  known analytically \cite{Assaf:2010eb,Mobilia:2010fk}
but are unnecessary for
illustrating the failure of the fixed population size model. Note that 
Eq. (\ref{eqn:plateau_height}) resembles the
Boltzmann weight in equilibrium statistical mechanics if $N$ is interpreted
as inverse temperature while $S_0[\gamma_{f^{*} \rightarrow x}]$  and $\ln
w[\gamma_{f^{*} \rightarrow 0} ]$ play the role of energy and entropy, respectively,
as in the classical
 Kramers escape-over-a-barrier problem  due to thermal fluctuations \cite{Risken:1984df}.
For $f^* >1/2
,$ it is more likely for species 1 to be fixed and we can infer from Eq.
(\ref{eqn:plateau_height})
that $\Delta S_0 > 0$, resulting in $u(f^*) \rightarrow 1$ as $N \rightarrow
\infty.$ Similar arguments give  $\Delta S_0 < 0 $ if  $f^* <1 /2,$ implying
that $u(f^*) \rightarrow 0$ as $N \rightarrow \infty.$  

\begin{figure}[th!]
\includegraphics[width = 0.9\linewidth]{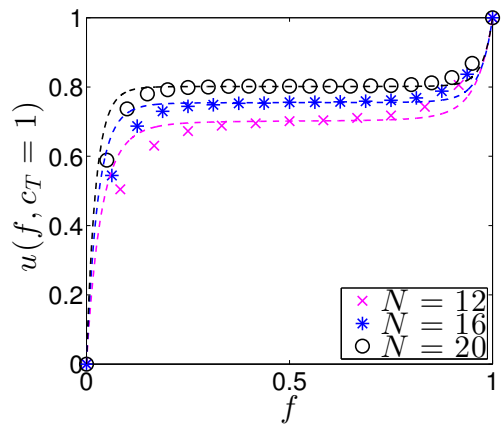}
\caption{\label{fig: Kimura_FixProb}(Color online) Comparison between fixed
population size predictions 
given by Eq. (\ref{eqn: FP_MutualismCrowded}) 
(dashed lines) and the simulated fixation probabilities from the Gillespie's
algorithm (symbols) for  an
 asymmetric strong mutualism with a varying population size with  $s_{o}=0,
\  \beta_1= 0.75,\ \textrm{and} \ \beta_2 = 0.70$. The plateau-fitting procedure
yields the fitting parameters $f^*_{eff} = f^*=\beta_1/(\beta_1+\beta_2)
\approx 0.517$ and  
$\tilde  \beta N_{eff}$ $= 58.0, \ 74.4,$ 
$91.0$ for the simulated $N = 12, \ 16, \ 20,$ respectively. This
procedure always  fits  the plateau, but fails to capture the boundary layer
behavior.}
\end{figure}

We now denote the two free parameters of Eq. (\ref{eqn: FP_MutualismCrowded})
by $\tilde \beta N_{eff}$ and $f^*_{eff}$
, and fit $u(f)$ to our numerically simulated fixation probability for  
 strong asymmetric mutualism with    $s_o=0, \ \beta_1 = 0.75,
\ \textrm{and}\ \beta_2 = 0.70 $ whose actual coexistence fixed point is
$(f^{*},c_T^*) \approx (0.517, 1.568). $  Our stochastic
simulations  employ the Gillespie algorithm to efficiently simulate the
discrete Master equation of Sec. \ref{subsec: (2C)stochastic} \cite{Gillespie:1976km,Gillespie:1977fk}.
The simulated fixation probabilities  for each
initial condition
are constructed from $10^4$ realizations of fixation events.
The initial overall population size
in our simulations is taken to be $c_T = 1 <c_T^*,$ i.e., the initial overall
population size is less than the fixed point value.  The simulated results
shown in Fig.  \ref{fig: Kimura_FixProb} reveal a plateau of equal
fixation probability  even for  relatively small $N \gtrsim\ 12$. To match
the center of the plateau,
we choose  $f^*_{eff}$ = $f^*$. The other free parameter
$\tilde \beta N_{eff}$  controls both the plateau height and width. Because
the plateau structure occupies most region, it is reasonable to adjust  $\tilde
\beta N_{eff}$
so that $u(f^*) $ matches the height of the simulated plateau.
With this fitting procedure, the plateau in the fixed population size
model is guaranteed to agree with the simulated plateau.

Although the fixed population size model can be adjusted to fit the elongated
plateau
in agreement with simulations,  it fails to capture the behavior near
the absorbing boundaries   as revealed by Fig.  \ref{fig:
Kimura_FixProb}, where the simulated points systematically fall away from
the predicted dashed lines.  In fact, it is precisely this  boundary behavior
that distinguishes
the fixation probability of mutualism with a fixed population size from mutualism
with a varying population size. As we now show, the elongated plateaus
also exist for strong mutualism with a varying population size,
but the behavior near absorbing boundaries depends on the delicate interplay
between the relative frequency and the overall population
size.

\subsection{The fixation probability and the mean fixation time from matched
asymptotic expansions}
\label{subsec: 5B(FP_MFT_Mutualism)}
\begin{figure}[ht!]
\includegraphics[scale = 0.35]{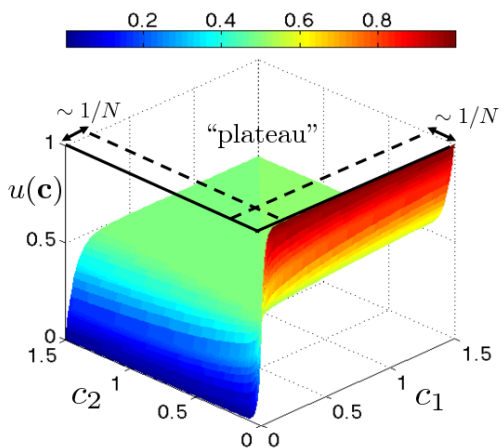}
\caption{\label{fig: bedprofile}(Color online) The  fixation probability
for  strong symmetric
mutualism with vanishing reproductive advantage near the origin ($s_{o}=0$),
 $\beta_1=\beta_2=0.75$  
and $N=20$ from the matched asymptotic expansions. In the boundary layers
adjacent to the absorbing boundaries, the fixation probability shows  crossovers
with the characteristic width $1/N$ from the boundary condition values to
the plateau value $P=1/2. $ Within the boundary layers, fixation at
early times before approaching the vicinity of the coexistence fixed point
is likely. In contrast, initial conditions in the plateau almost surely arrive
at
the coexistence fixed point before rare fluctuations eventually  leads to
fixation. The heat map provides an alternative representation of $u(\boldsymbol
c).$} 
\end{figure}
We now study the fixation probability  and the mean fixation time, taking
into account  
\textit{both} the frequency and the  population size degrees of freedom.
Our results 
for the fixation probability are summarized in Fig. \ref{fig: bedprofile}.
A fixation event
with  initial frequency $f$ and  initial population size $c_T$ requires
 a two-dimensional escape  to an absorbing boundary   from
the initial condition which we specify as $(fc_T,(1-f)c_T) $ in the $(c_1,c_2)
$ coordinates.
In contrast to mutualism under the replicator condition ($0<\beta_i \ll 1$
and $c_T \approx 1$), there is no dimensional
reduction to an effectively one-dimensional dynamics with approximately fixed
$c_T\approx1$  here. In fact, the fixation probability $u(\boldsymbol c)$
obeys a two-dimensional
backward Kolmogorov equation, namely 
\begin{equation}
0 = \sum_{i=1}^2\Big(v_i(\boldsymbol{c})\partial_{c_i}u(\boldsymbol{c})
+ \frac{1}{2N}D_i(\boldsymbol{c})\partial_{c_i}^2u(\boldsymbol{c}) \Big),
\label{eqn: BKE_2d}
\end{equation}     
with the deterministic drifts  $v_i(\boldsymbol
c)$ and diffusion coefficients  $D_i(\boldsymbol
c)$ given by Eqs. (\ref{eqn: drift1})-(\ref{eqn: diff2}). The absorbing boundaries
corresponding to the fixation of species 1 and of species 2 impose the
boundary conditions $u(c_{1},0)=1$ and $u(0,c_{2})=0,$ respectively. 
Eq. (\ref{eqn: BKE_2d}) does not admit an exact solution, and (as mentioned
above) the standard
technique of escape from a potential well can not be applied since $v_i(\boldsymbol
c)$ is  not a gradient of a potential function, i.e., $|\vec\nabla \times
\vec{v}(\boldsymbol c)|= |\mu_1(1-\beta_1)c_1-\mu_2(1-\beta_2)c_2|\neq  0.$
Despite these complications, given an empirical
data set
with a plateau structure of fixation probability
\textit{a priori}, we
can solve for $u(\boldsymbol
c)$ accurate to first order in $1/N$ by the method of  matched asymptotic
expansions \cite{Grasman:1999kx,Verhulst:2006fk,Bender:1999uq,Hinch:xy}.
The strategy is to separately find asymptotic solutions of
$u(\boldsymbol c)$ in the plateau region and in the boundary layers adjacent
to  the absorbing boundaries, and then perform  asymptotic matching of the
local solutions. Note that our analysis  follows from the Fokker-Planck approximation
to the Master equation.
It has been shown for initial conditions \textit{starting
from a metastable state}, for example in Refs. \cite{Ovaskainen:2010bc,Assaf:2010xp,Kessler:2007bj},
that
the quasi-stationary distribution (QSD) and the mean fixation time when fixations
occur via rare fluctuations  are accurately predicted  by the WKB approximation
of the  Master equation, rather than by the WKB approximation  of the Fokker-Planck
approximation. However, the  functional form of the QSD and of the mean fixation
time from the two methods coincide, and are given by Eq. (\ref{eqn: WKB})
and Eq. (\ref{eqn: plateau_MFT}) respectively. We show here that treating
the \textit{N}-independent parameters  in the functional form as fitting
parameters yields excellent  fits to the plateau fixation probability  and
the plateau mean fixation time caused by escape from a QSD. The utility of
the Fokker-Planck approximation  here is its ability to predict   \textit{crossover
behaviors} from the boundary values to the plateau values of the fixation
probability and the mean fixation time  from asymptotic expansions in $1/N.$
As  discussed in  Sec. \ref{subsec: 5A(mutual_fixprob)}, these crossovers
 are the essential feature of  strong mutualism with  varying population
sizes and, to the best of our knowledge, have not been calculated previously
by any technique. With the crossover behaviors in mind, we proceed with the
usual Fokker-Planck approximation of the Master equation.

In the plateau region, the fixation probability $u(\boldsymbol c)$ near the
coexistence fixed point
  $\boldsymbol c^* =(c_1^*,c_2^*) $ is approximately equal to $P\equiv u(\boldsymbol
c^*).$
Similar to  strong mutualism with a fixed
population size, the dynamics in the plateau region can be characterized
by a rapid
approach to the coexistence fixed
point  $\boldsymbol c^*  $ before weak fluctuations eventually drive
the system toward fixation by a large deviation.  Eq. (\ref{eqn:
BKE_2d}) guarantees the existence of the plateau structure if number fluctuations
are sufficiently weak. Indeed, in the limit $1/N \rightarrow 0,$   Eq. (\ref{eqn:
BKE_2d}) reduces to the simple advection equation  $0 = \sum_{i=1}^2v_i(\boldsymbol{c})\partial_{c_i}u(\boldsymbol{c}).$
The associated characteristics $\boldsymbol c(t)$ obey the mean field
dynamics $dc_i(t)/dt = v_i(\boldsymbol c)  $ on which $du(\boldsymbol
c(t))/dt =0 $, meaning that the fixation probability along each characteristic
is constant. Because all the 
characteristics 
  meet at the stable fixed point $   \boldsymbol c^*$, we conclude $u(\boldsymbol
 c) = P \equiv u(\boldsymbol c^*).$ This plateau value, however, can not
extend over the entire
domain
without violating  the boundary conditions; hence,  boundary layers adjacent
to the absorbing boundaries are an essential part of the physics. 

\begin{figure}[th!]
\subfigure{\includegraphics[width=\linewidth]{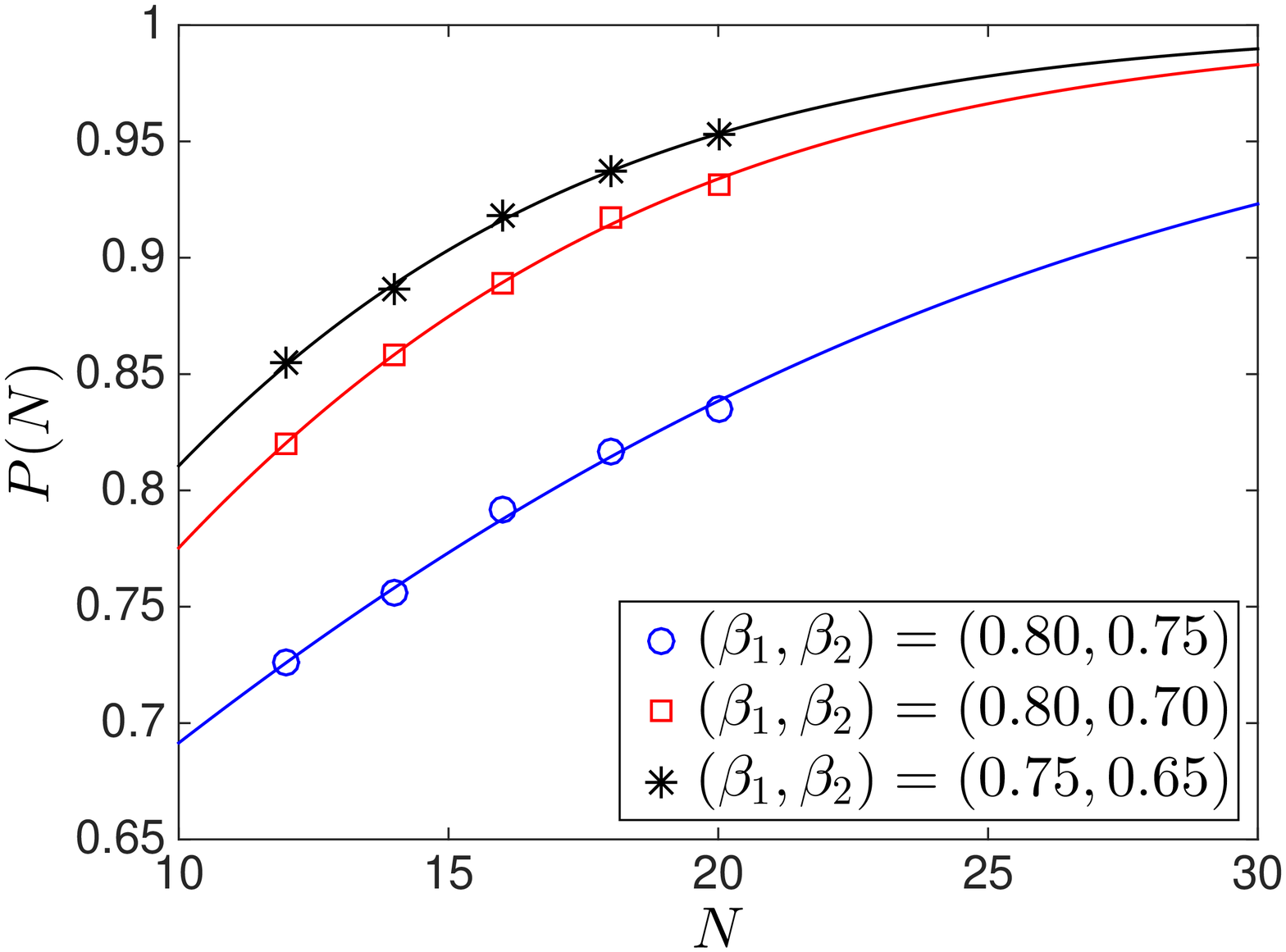}}
\subfigure{\includegraphics[width=\linewidth]{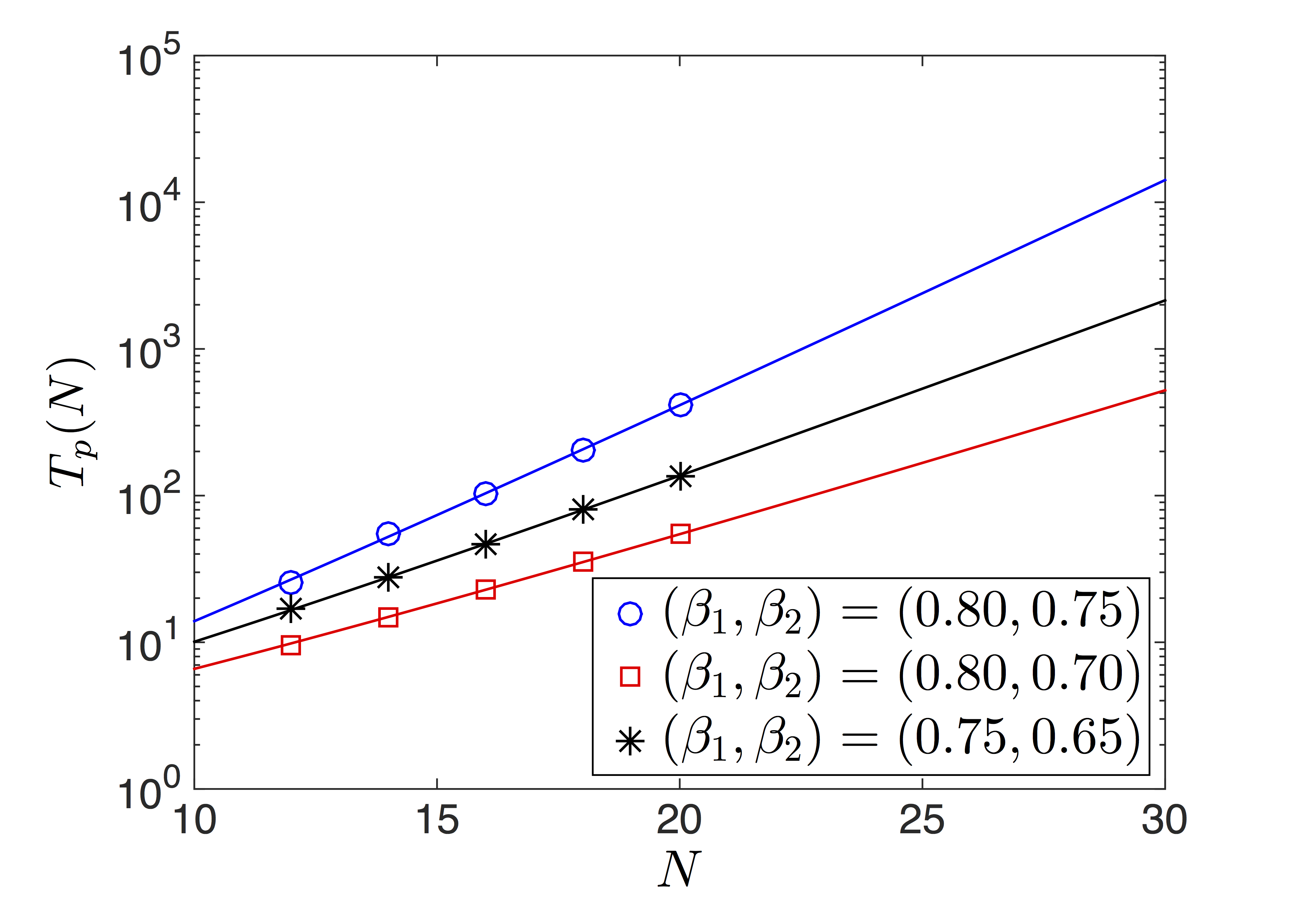}}
\caption{\label{fig: Fit_Plateau_FPMFT}(Color online) Extrapolations of the
plateau fixation probability
$P$ (top) based on Eq. (\ref{eqn:plateau_height_2d}), and a semilog plot
of the plateau mean
fixation time $T_{p}$ (bottom)
based on Eq. (\ref{eqn: plateau_MFT}), by  best fits to simulations  with
$N=12,14,16,18,20$ in different cases of slightly asymmetric strong mutualism.
Symbols are
simulation results and solid lines are best fitted curves based on 
Eqs. (\ref{eqn:plateau_height_2d}) and (\ref{eqn: plateau_MFT}). The coexistence
fixed point lies closer to the fixation of species 1, yielding $\Delta
S_0 > 0. $  The plateau fixation probability $P$ increases at increasing
$N$ and saturates at 1 as $N \rightarrow \infty.$ Note that the plateau mean
fixation
time $T_{p}$  grows exponentially with $N$.} 
\end{figure}

Analogous to $u(f^*)$ in the previous subsection, the plateau fixation probability
$ P \equiv u(\boldsymbol c^*)$ is the ratio of the
flux into the absorbing boundary  $f=1 $ to  the total flux into both of
the absorbing
boundaries  $f=0$ and $f=1$, see Appendix \ref{sec:plateau_QSD}. The probability
flux
 peaks at the saddle
fixed point of each absorbing boundary, which
suggests that these saddle fixed points dominate  the most probable escape
routes for 
each absorbing boundary. Appendix \ref{sec:plateau_QSD} discusses the derivation
of  the $N$-dependence of $P  $ which takes the asymptotic form similar to
Eq. (\ref{eqn:plateau_height}):
\begin{equation}
P \approx \frac{1}{1+e^{-N\Delta S_0 + \Delta S_1}},\label{eqn:plateau_height_2d}
\end{equation}
where  $\Delta S_0$ and $\Delta S_1$ are treated here as fitting parameters.
For   strong symmetric mutualism with $s_{o}=0$ and $\beta_1 = \beta_2,$
the
dynamics has a reflection symmetry with respect to the line   $f = 1/2
$ in the $c_1$-$c_2$ plane; hence, fixation of either species is equally
likely and $P = 1/2$ independent of $N$. In the symmetric case, we can
thus infer $\Delta S_0 = \Delta S_1 = 0$. For strong asymmetric mutualism,
we expect that, in the limit $N\gg 1,$ species 1 is more likely to be fixed
 if  $f^*>1/2$  , and hence $\Delta S_0 > 0$. This assertion is
confirmed by simulations in Fig. \ref{fig: Fit_Plateau_FPMFT} where  
 we find $\Delta S_0 > 0 $ for  $f^*>1/2$ from fitting to  Eq. (\ref{eqn:plateau_height_2d}).
Extrapolations of
 fits to Eq. (\ref{eqn:plateau_height_2d}) imply that if $f^* > 1/2,$
then species 1 will be fixed with probability 1 in the limit $N \rightarrow
\infty.$ 
       
In the boundary layers (see Fig. \ref{fig: bedprofile}),
  $u(\boldsymbol c)$ crosses over  from
the boundary condition values to the plateau value. This crossover embodies
two
different types of dynamics: fixation
at early times without falling into the basin of attraction of   $\boldsymbol
c^*$  
and  fixation by a large deviation after captured by   $\boldsymbol c^*$.
Fixation at
early times is possible if the initial condition lies within the boundary
layers, whose characteristic width is    $1/N \ll 1. $ The characteristic
widths
and the behavior of crossovers  can be extracted by the method of matched
asymptotic expansions \cite{Grasman:1999kx,Verhulst:2006fk,Hinch:xy,Bender:1999uq}.
The details of matched asymptotic expansions are
given by Appendix \ref{sec:MAE}.  
\begin{figure}[th!]
\subfigure{\includegraphics[width=\linewidth]{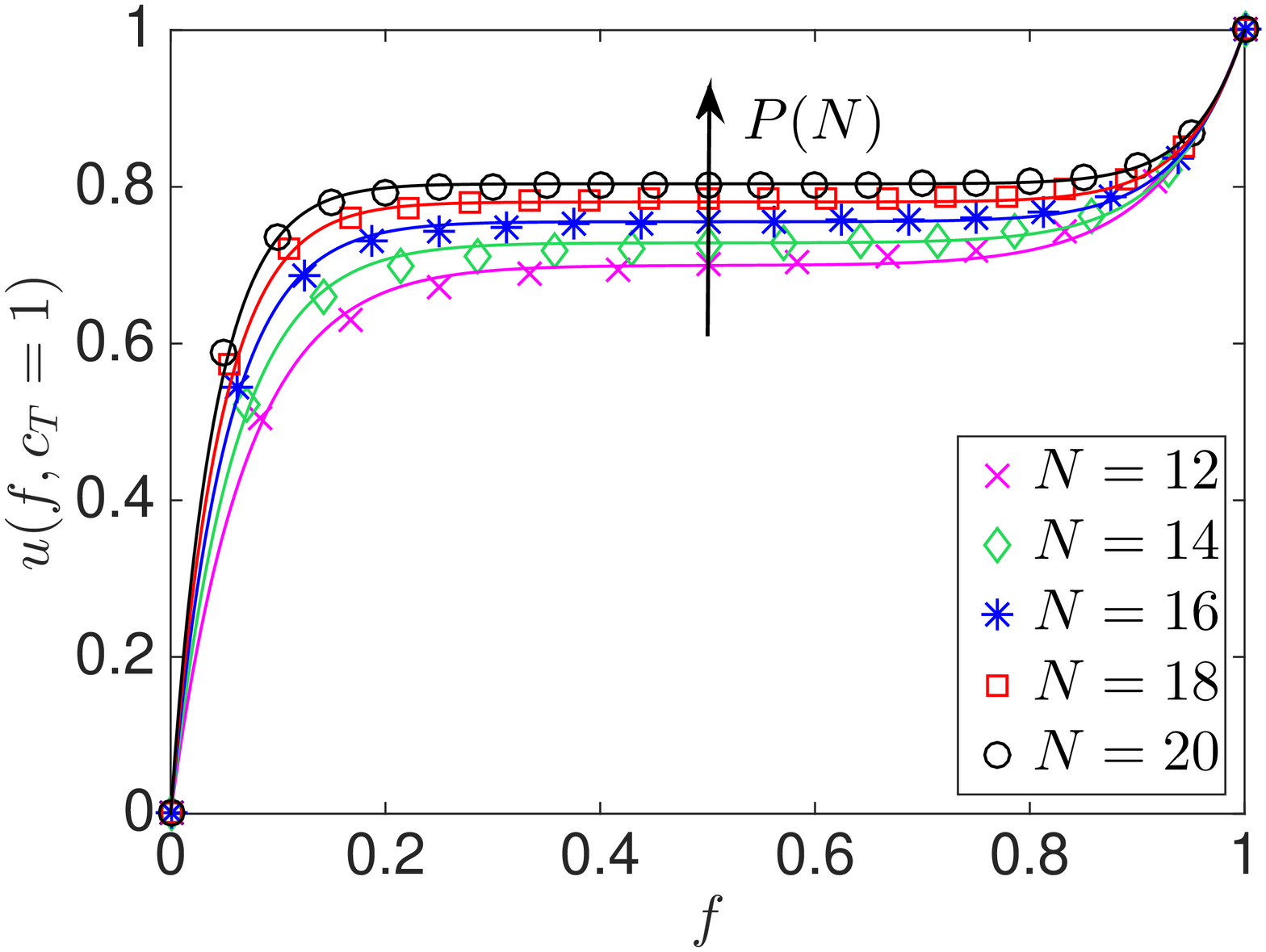}}
\subfigure{\includegraphics[width=\linewidth]{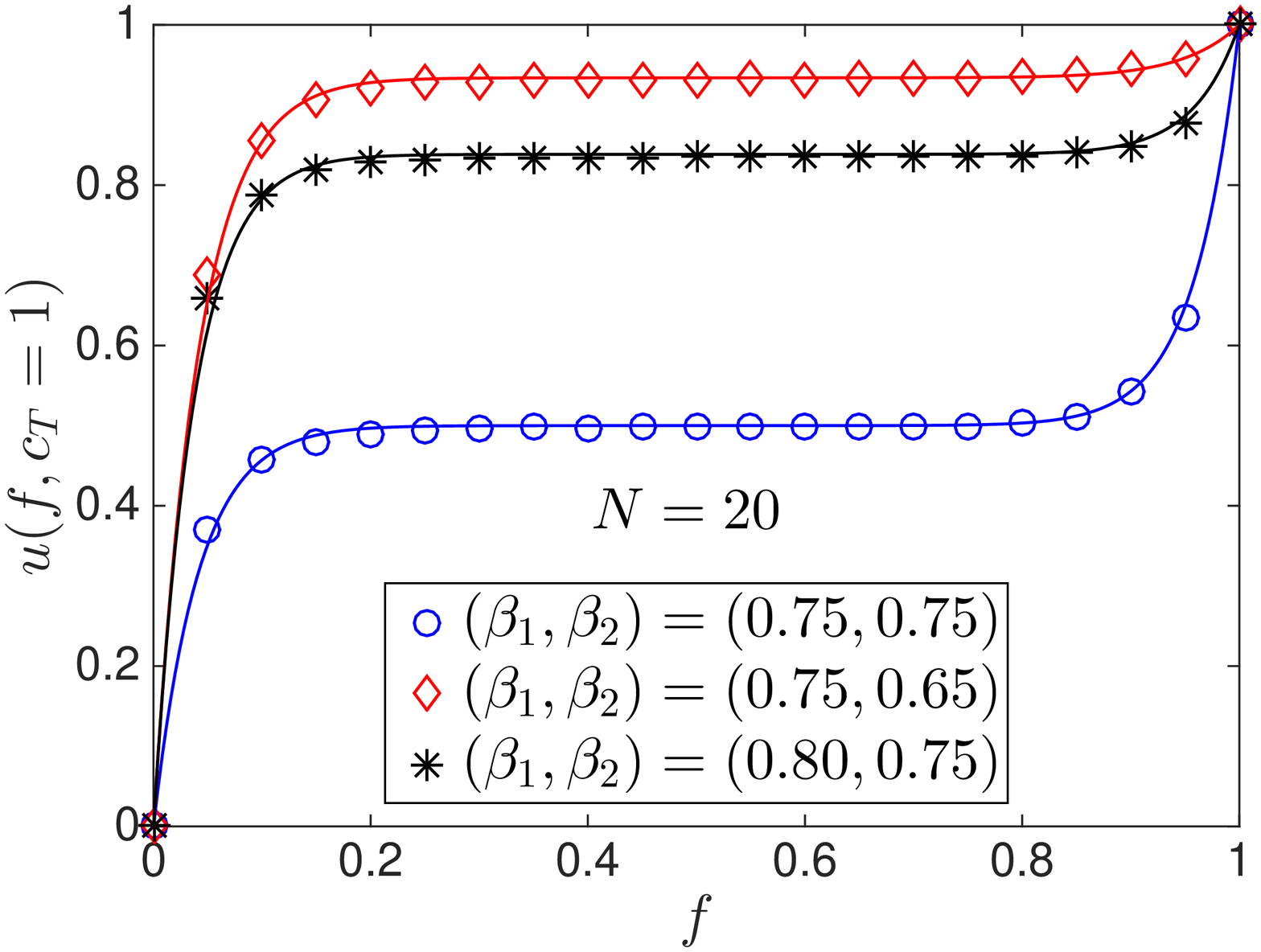}}
\caption{\label{fig: FP_MatchedAsymp_V_Sim}(Color online) Comparisons between
the 
predicted fixation probability
(solid lines) and simulation results (symbols). The plateau fixation probabilities
$P$ are determined by fitting Eq. (\ref{eqn:plateau_height_2d}) to simulations
with $N=12,14,16,18,20$ with other parameters fixed. Top: the fixation probability
for $s_o=0, \beta_1 = 0.75, \beta_2=0.70$ 
at increasing $N$, indicated by the arrow. The matched asymptotics yield
excellent estimates
for the crossovers from the boundary values to the plateau value $P(N)$
even for $N$ as small as 12. As $N$ increases, the plateau region  becomes
more elongated (i.e., the boundary layers have characteristic width $1/N$)
while the plateau
value  $P(N)$ increases and eventually saturates at 1 as $N \rightarrow \infty,$
similar to Fig. \ref{fig: Fit_Plateau_FPMFT}(top). Bottom: the fixation probability
at $N=20$ for
different cases of strong mutualism. Note the improved agreement between
simulation and theory compared to  Fig. \ref{fig: Kimura_FixProb}.} 
\end{figure}

In the boundary layer adjacent to the absorbing boundary $c_2=0$ and away
 the absorbing boundary $c_1 =0$,
the asymptotic large $N$ form of the fixation probability reads
\begin{equation}
u(\boldsymbol c) = P +(1-P)  e^{-N c_2 /\Phi_1(c_1)},\label{eqn: Main_AsymptoticC1Fixed}
\end{equation}
where the function $\Phi_{1}(x)$ satisfies 
\begin{eqnarray}
0 = &&-x(1-x)\Phi_1'(x)+ (1+s_o)^{-1}
[1 - (1-\beta_2)x]\Phi_1(x)\nonumber\\
&&-\frac{1}{2}(1+s_{o})^{-1}
[1 + (1-\beta_2)x], \label{eqn: main_phi1}
\end{eqnarray}
with the matching condition $\lim_{x \rightarrow 1} \Phi_1(x) = \frac{2-\beta_2}{2\beta_2}.$
For a fixed $c_1$, Eq. (\ref{eqn: Main_AsymptoticC1Fixed}) implies that the
fixation
probability exhibits a crossover from 1 to $P$\ as $c_2$ increases from
$c_2=0$ to $c_{2} \gg 1/N.$  The details of  the crossover depend on  $1/\Phi_1(x),
$ which  is  a monotonically decreasing function of $x$ for $\beta_1 < 1$
and
$\beta_2 < 1,$ with $1/\Phi_1(0)=2.$    

In the complementary boundary layer
adjacent to the absorbing boundary $c_{1}=0$ and away from  the absorbing
boundary $c_2 =0$,
the asymptotic form of the fixation probability reads  
\begin{equation}
u(\boldsymbol c) = P -P  e^{-N c_1 /\Phi_2(c_2)},\label{eqn: Main_AsymptoticC2Fixed}
\end{equation}
where the function $\Phi_{2}(x)$ satisfies 
\begin{eqnarray}
0 = &&-x(1-x)\Phi_2'(x)+ (1+s_{o})
[1 - (1-\beta_1)x]\Phi_2(x)\nonumber\\
&&-\frac{1}{2}(1+s_{o})
[1 + (1-\beta_1)x], \label{eqn: main_phi2}
\end{eqnarray}
with the matching condition $\lim_{x \rightarrow 1} \Phi_2(x) = \frac{2-\beta_1}{2\beta_1}.
  $ For a fixed $c_2$, Eq. (\ref{eqn: Main_AsymptoticC2Fixed}) implies that
the
fixation
probability exhibits a crossover from 0 to $P$\ as  $c_1$ increases 
from $c_{1}=0$ to $c_1 \gg 1/N.$ Similar to $1/\Phi_1(x),$  $1/\Phi_2(x)$
is a monotonically decreasing  function of $x$ in the parameter range of
interest with  $1/\Phi_2(0)=2.$

The  above local behaviors of fixation probability can be combined into
the global asymptotic solution  
\begin{equation}
u(\boldsymbol c) = P + (1-P)e^{-Nc_{2}/\Phi_1(c_{1})} - Pe^{-Nc_{1}/\Phi_2(c_{2})},
\label{eqn: FixProb_Mutualism_c1c2}
\end{equation}
where the functions $\Phi_1(c_{1})$ and $\Phi_2(c_{2})$ obey Eq. (\ref{eqn:
main_phi1}) and Eq. (\ref{eqn: main_phi2}) with the associated matching
conditions. This global asymptotic solution valid everywhere on the domain
except in
the small box near the origin $[0,1/N]
\times [0,1/N] $ where the two boundary layers overlap. Upon changing the
coordinate to $(f,c_T)$ to emphasize the important population size
degree of freedom, we obtain, finally, 
\begin{eqnarray}
u(f,c_T) = P &&+ (1-P)e^{-N(1-f)c_{T} /\Phi_1(fc_{T})} \nonumber\\
&&- Pe^{-Nfc_{T}/\Phi_2((1-f)c_{T})}.
\label{eqn: FixProb_Mutualism_fct}
\end{eqnarray}
Fig. \ref{fig: bedprofile} summarizes the fixation probability
as a function of $(c_1,c_2)$ predicted by Eq. (\ref{eqn: FixProb_Mutualism_c1c2})
for a strong symmetric mutualism. Fig. \ref{fig: FP_MatchedAsymp_V_Sim}
shows excellent agreement between  the prediction of Eq. (\ref{eqn: FixProb_Mutualism_c1c2})
and the simulation results for a strong asymmetric mutualism. As expected,
the improvements relative to Fig. \ref{fig: Kimura_FixProb} occur for $f$
near 0 and 1:  the boundary
behavior missed by the  fixed population size model  are now well captured
even for $N$ as small as 12.
\begin{figure}[th!]
\subfigure{\includegraphics[width=\linewidth]{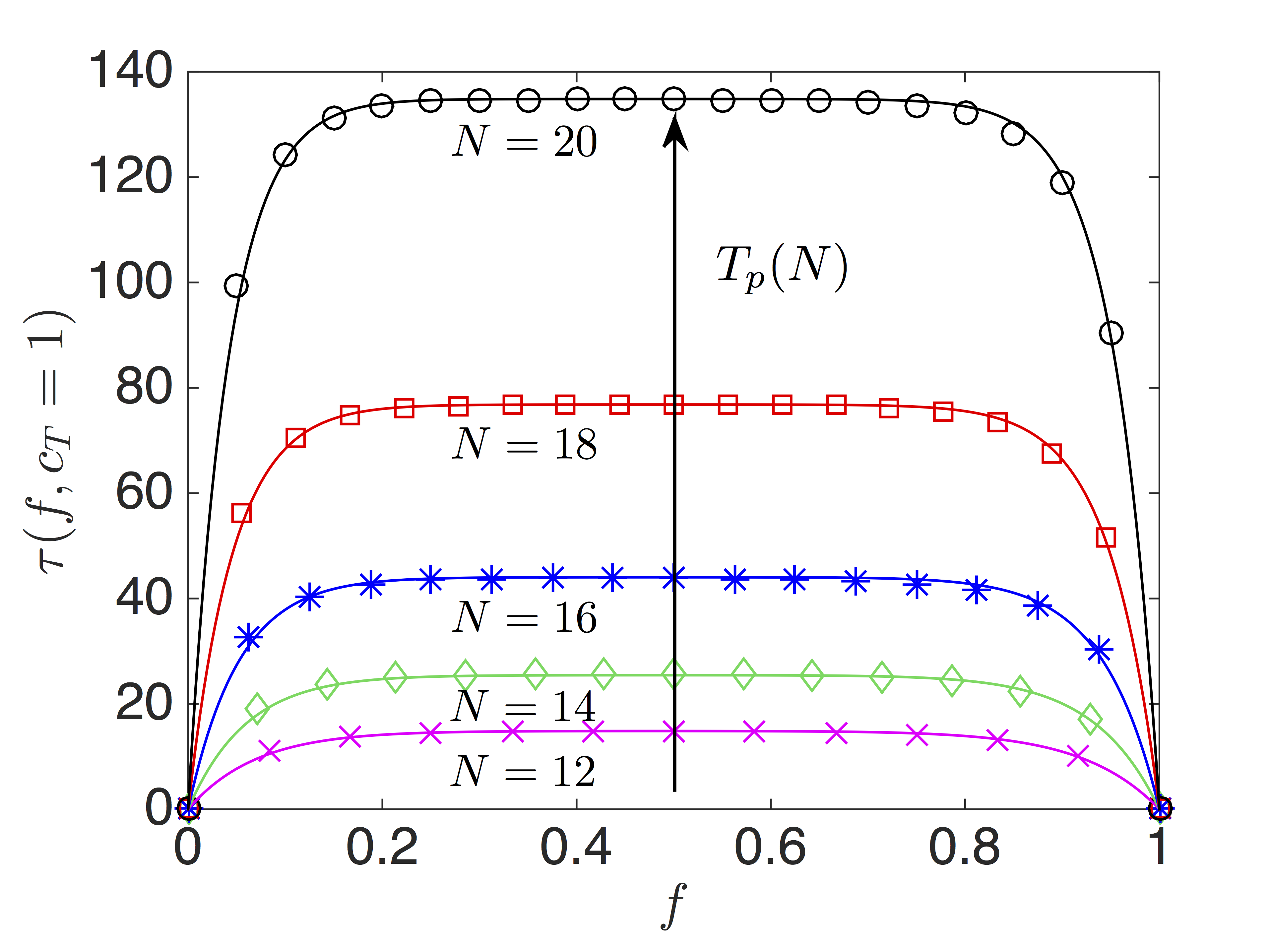}}
\subfigure{\includegraphics[width=\linewidth]{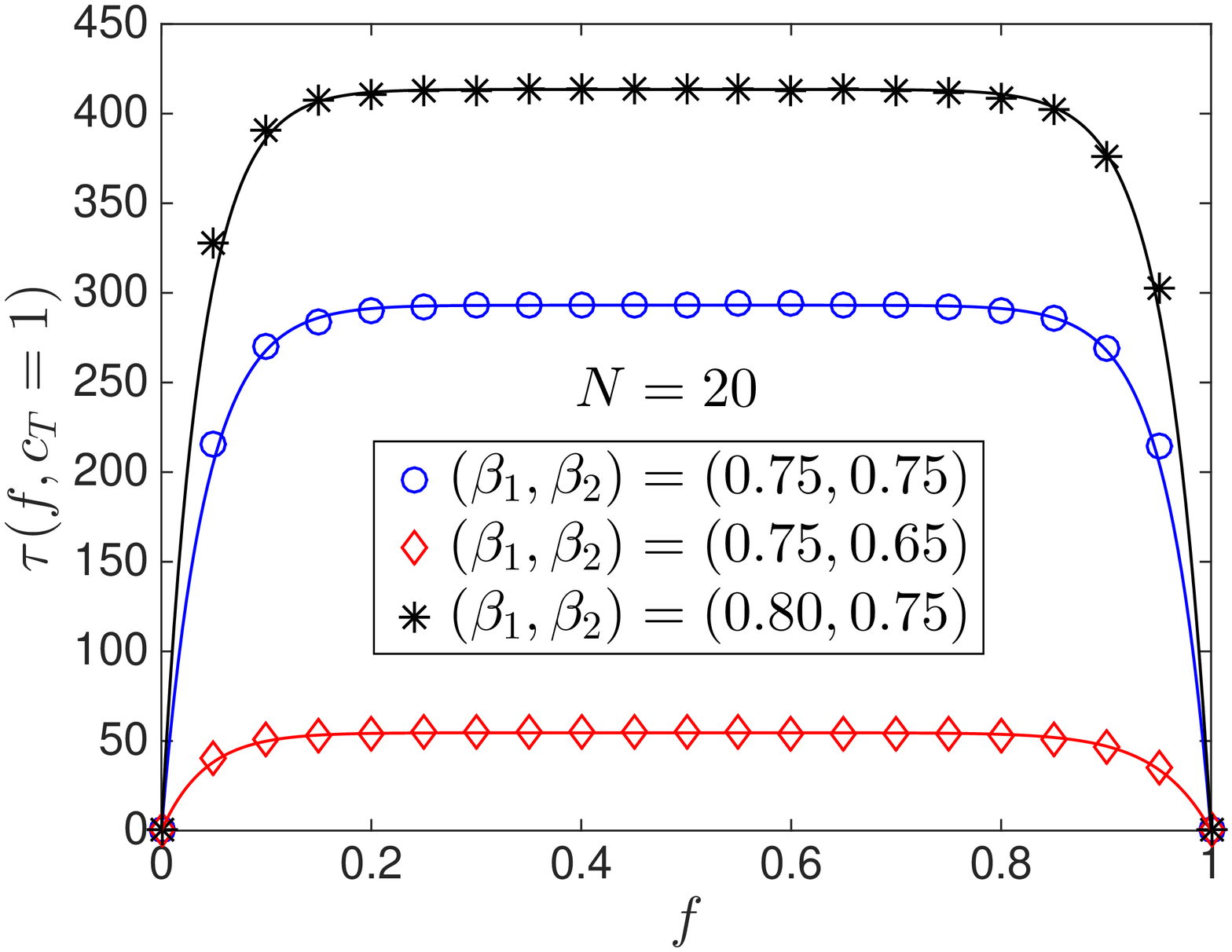}}
\caption{\label{fig: MFT_MatchedAsymp_V_Sim}(Color online). Comparisons 
between the predicted mean fixation time
(solid lines) and simulation results (symbols). Parameters are the same
as  Fig. \ref{fig: FP_MatchedAsymp_V_Sim}'s. The plateau mean fixation times
are determined by fitting, Eq. (\ref{eqn: plateau_MFT}), to simulations with
$N=12,14,16,18,20$
with other parameters fixed. Top: the mean fixation time for $s_{o}=0, \beta_1
= 0.75, \beta_2=0.70$ 
at increasing $N.$ The matched asymptotic expansions yield excellent
estimates for the crossovers from the boundary values to the plateau  value
$T_{p}(N)$ even
for $N$ as small as 12. As $N$ increases, the plateau region  becomes more
elongated (i.e., the boundary layers have characteristic width $1/N$) while
 the plateau
value $T_{p}(N)$ grows exponentially similar to Fig. \ref{fig: Fit_Plateau_FPMFT}(bottom).
Bottom: the mean fixation time at
$N=20$ for different cases of strong mutualism.} 
\end{figure}

The mean fixation time, $\tau(\boldsymbol c)$, can also be constructed by
the method of matched asymptotic expansions. 
In this case, we need to solve \cite{Gardiner:1985qr,Van-Kampen:1992tk}
\begin{equation}
-1= \sum_{i=1}^2\Big(v_i(\boldsymbol{c})\partial_{c_i}\tau(\boldsymbol{c})
+ \frac{1}{2N}D_i(\boldsymbol{c})\partial_{c_i}^2\tau(\boldsymbol{c}) \Big),
\label{eqn: BKE_MFT_2d}
\end{equation}
subject to the boundary conditions $\tau(c_{1},0)=0$ and $\tau(0,c_{2})=0.
  $ Asymptotic matching arguments similar to Appendix \ref{sec:MAE} can be
applied
to Eq. (\ref{eqn: BKE_MFT_2d}), resulting in the  global asymptotic solution
for the mean fixation time, namely 
\begin{equation}
\tau(\boldsymbol c) =T_{p}
 \left(1-e^{-Nc_{2}/\Phi_1(c_{1})} - e^{-Nc_{1}/\Phi_2(c_{2})}\right),
\label{eqn: MFT_2d}
\end{equation} 
where the functions $\Phi_1(c_{1})$ and $\Phi_2(c_{2})$ still obey Eq.
(\ref{eqn:
main_phi1}) and Eq. (\ref{eqn: main_phi2}), and $T_{p}$ is the plateau mean
fixation
time for the initial
condition at the coexistence fixed point. The function $\tau(\boldsymbol
c)$ possesses a plateau structure in which   $\tau( \boldsymbol c) \approx
\tau(\boldsymbol c^*) = T_{p},$ similar to the profile of  $u(\boldsymbol
c)$. Furthermore, crossovers from the boundary
conditions to    $T_{p}$  are  characterized by the same exponentials  $e^{-Nc_{2}/\Phi_1(c_{1})}$
 and $e^{-Nc_{1}/\Phi_2(c_{2})}$ as in Eq. (\ref{eqn: FixProb_Mutualism_c1c2}).
In the limit $N \gg1$, the behavior of  $T_{p}$ is exponential in $N$,  
\begin{equation}
T_{p} \approx \sigma_{0} e^{N\sigma_1}, 
\label{eqn: plateau_MFT}
\end{equation} 
where we treat $\sigma_0$ and $\sigma_1$ as fitting parameters   
 \cite{Elgart:2004ya,Kamenev:2008xu,Grasman:1999kx}.
Fig.  \ref{fig: Fit_Plateau_FPMFT}(bottom) confirms the exponential scaling
of
Eq. (\ref{eqn: plateau_MFT}), while Fig. \ref{fig: MFT_MatchedAsymp_V_Sim}
reveals excellent agreement between   Eq. (\ref{eqn: MFT_2d})
and the simulation results. To again emphasize the importance of the population
size degree of freedom, we rewrite Eq. (\ref{eqn: MFT_2d})
in the coordinates $(f,c_T)$ as
\begin{eqnarray}
\tau(f,c_T) =T_{p}
 \Big(1&&-e^{-N(1-f)c_{T}/\Phi_1(fc_T)}\nonumber\\ 
 &&- e^{-Nfc_{T}/\Phi_2((1-f)c_T)}\Big).\label{eqn: MFT_2d_fct}
\end{eqnarray}

\section{conclusions}
\label{sec:conclusion}
We have explored the interplay between   evolutionary dynamics
and population dynamics in a well-mixed competitive Lotka-Volterra model
in various limits. 
 The model gives rise to    5 different  scenarios, similar to
evolutionary game theory,  without  however fixing the overall population
size, thereby demonstrating
an explicit microscopic system exhibiting  the feedback between evolutionary
dynamics and population dynamics phenomenologically studied in
Refs. \cite{Melbinger:2010wq,Cremer:2011pi}.  

In the limit $|\beta_1| \ll1$, $|\beta_2| \ll 1, $ and $1/N \ll 1$, with
an
arbitrary reproductive advantage near the origin $s_o$,
describes  rapid  relaxational dynamics of population size toward a \textit{fixed}
equilibrium  size along
a quasi-deterministic growth trajectory on which  $\rho \equiv
(1-f)f^{-1/(1+s_o)}c_T^{s_o/(1+s_o)}$ is constant. The  variable $\rho$ 
relates  the  population frequency $f$ to  the total
population size  $c_T$ as $c_T$ approaches the quasi-equilibrium at $c_T
\approx
1$:
The  frequency of a reproductively advantageous species, on average, 
increases as  dilute populations ($c_T <1$)  grow, and  decreases as overcrowded
populations ($c_T>1$)  decline. For $s_o=0$,  replicator dynamics
with genetic drift is recovered when $c_T \approx 1,$   despite population
size fluctuations away from $c_T =1$.
Only in this limit is the dynamics near the equilibrium population size a
simple generalization  of conventional
population genetics without mutation with an independently fluctuating
 population size. From the perspective of equilibrium statistical mechanics,
this simple limit is analogous to  the generalization from a canonical ensemble
to a
grand canonical ensemble in the thermodynamic limit \cite{Pigolotti:2013le}.
However, for $s_{o}
\neq 0$, population size fluctuations couple to evolutionary dynamics in
a nontrivial fashion and replicator dynamics with genetic drift is no longer
an appropriate description. Our
results demonstrate explicitly a circumstance such that the fixed effective
population size
model in population genetics is incomplete.

It would be interesting to study how population size fluctuations
affect evolutionary dynamics near a fixed equilibrium population size when
$s_{o} \neq 0$ in
selection scenarios other than  quasi-neutral evolution studied in Refs.
\cite{Parsons:2007aa,Parsons:2008aa,Kogan:2014aa,Lin:2012wf}.  Particularly
interesting  is the
Prisoner's  Dilemma briefly
discussed at the end of Appendix
\ref{sec:appndxA}. In this case,  fluctuation-induced selection can actually
oppose
the usual selection bias in the  Prisoner's Dilemma, illustrating a fluctuation-driven
mechanism other than genetic
drift  (or spatial segregation \cite{Dyken:2013ml})  that can alleviate the
dilemma of cooperation \cite{Cremer:2009sh}.

 We also studied competitions that take place with a strongly varying
population size (as opposed to competition with a nearly fixed  population
size), as in the strong mutualism limit.  Fixation events can now arise via
two distinct mechanisms: fixation at long times by rare escape from the strongly
attractive coexistence
fixed point, and fixation at early times before reaching the neighborhood
of the coexistence
fixed point. The former situation is typical for initial conditions
well away from the absorbing boundaries  $(c_1,0)$ and $(0,c_2)$, where the
system
initially falls toward
the coexistence fixed point, resulting in a plateau of constant
fixation probability and a plateau in the  mean fixation time. The latter
situation
arises for initial
conditions lying close to the absorbing boundaries where fluctuations
     can fix one species  before falling into the  coexistence fixed point.
 The crossovers
 from the absorbing boundaries to the plateau in the fixation probability
and the mean fixation time can be studied
by matched asymptotic expansions, accounting for number fluctuations,
evolutionary dynamics,
and population dynamics. As shown in  Fig.  \ref{fig:
Kimura_FixProb}, the fixed population size model underestimates the number
of fixation events that can occur near the absorbing boundaries, thereby
overestimating the probability and duration of species coexistence. This
dynamics can be important
in the context of range expansions of mutualists \cite{Lavrentovich:2014oj,Muller:2014wo}
where populations at the expanding frontier are continuously subject to interaction
in a growing population size, which may alter parameter values
separating an active (mutualistic) phase from inactive (single species
domination) phase.

Although our analytical predictions from the Fokker-Planck approximation
  show excellent agreement with numerical simulations of the Master equation,
it would be interesting to study strong mutualism from other approaches such
as the WKB approximation of the Master equation \cite{Ovaskainen:2010bc,Assaf:2010xp,Kessler:2007bj}.
These approaches can accurately predict the  quasi-stationary
distribution and the plateau mean fixation time  when
fixations  occur by large deviations from a metastable state without resorting
to fitting parameters of the plateau values.  Lastly, the
fate
of competitions as a function of both population size and the frequency
in other competition scenarios with a varying population size under strong
selection
 would also be worth investigating.
  
\begin{acknowledgments}
We thank Mogens Jensen, Maxim Lavrentovich, Simone Pigolotti, and Melanie
M\"{u}ller for helpful discussions. This work was supported in part by the
National Science Foundation (NSF) through grant DMR-1306367, and by the Harvard
Materials Research Science and Engineering Center through grant DMR-1420570.
Portions of this research were conducted during a stay at the Center for
Models of Life at the Niels Bohr Institute, the University of Copenhagen.
Computations were performed on the Odyssey cluster supported by the FAS Division
of Science, Research Computing Group at Harvard University. 
\end{acknowledgments} 

\appendix
\section{ Coupled dynamics of $f$ and $c_T$}
\label{sec:appndxA}
Upon applying Ito's change of variable to Eq. (\ref{eqn: langevin_ci}) and
denoting $\mu = \mu_2$ and $(1 + s_o)\mu = \mu_1$ \cite{Gardiner:1985qr,Van-Kampen:1992tk},
the coupled
 stochastic dynamics of 
$f$ and $c_T$ are described by   
\begin{align}
\frac{df}{dt} = \ &\mu v_R(f,c_{T}) +\mu \left(c_T + \frac{1}{N}\right)v_E(f)\nonumber\\
&+\sqrt{\frac{\mu D_{R}^{(f)}(f,c_{T})+\mu D_{E}^{(f)}(f)}{N}}\Gamma_f(t),\\
\frac{dc_T}{dt} = \  &\mu(1+s_{o}f)v_{G}(c_T) +\mu(\alpha_1 + \alpha_2)c_T^2
f(1-f)\nonumber \\ 
&+\sqrt{\frac{\mu D_{R}^{(c_{T})}(f,c_{T})+\mu D_{E}^{(c_T)}(f,c_{T})}{N}}\Gamma_{c_T}(t),
\end{align}
where  the $N$-independent functions in the deterministic drifts and in
the strength of an uncorrelated Gaussian white noise with $\langle \Gamma_a(t)\Gamma_b(t')\rangle
= \delta_{ab}\delta(t-t') $ and $\langle \Gamma_a(t)\rangle = 0$ are  given
by  
\begin{align}
v_R(f,c_{T}) &=  \left[(1-c_T) - \frac{1}{N}\left(\frac{1+c_T}{c_T} \right)
\right]s_{o} f(1-f),\\
v_E(f) &= f(1-f)[\alpha_1 +(\alpha_1 + \alpha_2)
f],\\
v_{G}(c_T) &=  c_T(1-c_T),\\
D_R^{(f)}(f,c_{T}) &=  f(1-f)[1+s_{o}(1-f)]\left( \frac{1+c_T}{c_T} \right),
\\
D_E^{(f)}(f) &=  - f(1-f)[\alpha_1(1-f)^2 + \alpha_2
f^2],\\
D_{R}^{(c_T)}(f,c_{T}) &=  c_{T}(1+c_T)(1+s_{o}f),\\
D_{E}^{(c_T)}(f,c_{T}) &= - (\alpha_1 + \alpha_2)c_{T}^2
f(1-f).
\end{align}
Here, the subscript $R$ denotes a contribution involving the reproductive
advantage
near the origin
$s_{o}$,  whereas the subscript $E$ denotes the contribution from evolutionary
parameters defined in Eq. (\ref{eqn: map_beta_alpha}) $ \alpha_1 = (1+s_{o})\beta_1
$ and $\alpha_2 = \beta_2$, and   $v_E(f)$
and $v_G(c_T)$ describe the deterministic replicator dynamics and the logistic
growth dynamics respectively. The $\mathcal{O}(1/N) $ contributions to  the
deterministic drift induced
by number fluctuations of $c_1$ and $c_2$ only appear in $df/dt$ and originate
from Ito's change of variable formula \cite{Gardiner:1985qr,Van-Kampen:1992tk}.
 
Under the replicator condition ( $|\alpha_1|
\ll 1$ , $|\alpha_2| \ll 1,$ and $c_T \approx 1$), we have  
$ \left|D_E^{(f)}/D_R^{(f)} \right| \ll 1 $ and $ \left|D_E^{(c_T)}/D_R^{(c_T)}
\right| \ll 1   $
so we can neglect the contributions from evolutionary parameters in the noise.
Therefore, at $_{}s_{o}=0$ the equations simplify,
\begin{align}
\frac{df}{dt} =\ &\mu v_E(f) \left(c_T + \frac{1}{N}\right)\nonumber\\
&+\sqrt{\frac{\mu}{N}f(1-f)\left(\frac{1+c_{T}}{c_T}
\right)}\Gamma_f(t), \label{eqn: appndx_dfdt_s0}\\
\frac{dc_T}{dt} =\ &\mu   v_{G}(c_T)+\mu \  (\beta_1 + \beta_2)c_T^2 f(1-f)\nonumber\\
&+ \sqrt{\frac{\mu}{N}c_T(1+c_T) }\Gamma_{c_T}(t).
\label{eqn: appndx_dcTdt_s0}
\end{align}
When $c_T \approx 1,$ (\ref{eqn: appndx_dfdt_s0}) reduces to Eq. (\ref{eqn:dfdt_s0})
and (\ref{eqn: appndx_dcTdt_s0})
reduces to Eq. (\ref{eqn:dcTdt_s0}) in the limit $|\beta_1 + \beta_2|\ll1/N
\ll 1. $
  
For $s_{o} \neq 0$   and $1/N \ll 1$, the coupled stochastic dynamics when
$c_T \approx 1$ acquires contributions from the reproductive advantage near
the origin $s_o$.
The dynamics is now described by   
\begin{align}
\frac{df}{dt} =\ &\mu v_R(f,c_{T}) +\mu c_{T}v_E(f)\nonumber\\
&+\sqrt{\frac{\mu}{N}f(1-f)[1+s_{o}(1-f)]\left(\frac{1+c_{T}}{c_T}
\right)}\Gamma_f(t), \label{eqn: appndx_dfdt_s}\\
\frac{dc_T}{dt} =\ &\mu(1+s_{o}f)v_{G}(c_T) +\mu(\alpha_1 + \alpha_2)c_T^2
f(1-f) \nonumber\\
&+\sqrt{\frac{\mu}{N}c_T(1+c_T)(1+s_{o}f)}\Gamma_{c_T}(t). \label{eqn:appndx_dcTdt_s}
\end{align}
In quasi-neutral evolution $(\alpha_1 = \alpha_2 = 0),$ the dynamics of $f$
on the equilibrium line $c_T = 1$ acquires the fluctuation-induced selection
term $v_R(f,c_T=1) = -2s_o f(1-f)/N,$ which actually favors the fixation
of the species with a reproductive \textit{disadvantage} near the origin
($s_0 <0$). The presence of non-vanishing deterministic drift is in stark
contrast to the unbiased random walk behavior of neutral evolution along
the equilibrium line displayed in Eq.(\ref{eqn:dfdt_neutral}). 

Another interesting limit also arises when $ \alpha_1= -\alpha_2 \sim \mathcal{O}(s_{o}/N),$
where
the fluctuation-induced selection term $\mu v_R(f,c_T = 1)=-(2\mu s_{o}/N)f(1-f)$
can compete
with the usual fixed population size selection strength $\mu v_E(f) = \mu
\alpha_1f(1-f)$ and the genetic
drift $D_R^{(f)}(f,c_T =1)=(2\mu/N)f(1-f)[1+s_{o}(1-f)]. $  In this case,
fluctuation-induced
selection can oppose the standard  selection in population genetics, or
equivalently in the Prisoner's Dilemma of evolutionary game theory, and thus
influence
the dilemma of cooperation.

\section{Matched asymptotics  for strong mutualism with a varying population
size}
\label{sec:MAE}

In this appendix, we construct the fixation probability for strong mutualism
from the method of matched asymptotic expansions, or equivalently the boundary
layer method, discussed in \cite{Grasman:1999kx,Bender:1999uq,Hinch:xy,Verhulst:2006fk}.
First, consider the asymptotic large $N$ solution $u(c_1,c_2)$ of the backward
Kolmogorov equation (\ref{eqn: BKE_2d}) near the saddle fixed points.
 In the neighborhood of the fixed point $(0,1),$ we introduce the stretched
coordinates
$\eta_1 = c_1N $ and $\eta_2 = (c_2-1)\sqrt{N }$ , and rewrite the fixation
probability
in the new coordinates as $U(\eta_1
,\eta_2)=u(\eta_1/N,1+\eta_2/\sqrt{N}).$ Upon neglecting the terms of
$\mathcal{O}(1/\sqrt{N})$,   Eq. (\ref{eqn: BKE_2d})
in the new coordinates reads
\begin{eqnarray}
0=\ &&\mu_1 \beta_1\eta_1\partial_{\eta_1}U + \Big(\frac{\mu_1+\mu_2}{2}-\frac{\mu_1
\beta_1}{2}\Big)
\eta_1\partial^2_{\eta_1}U\nonumber\\
&&-\mu_2\eta_2\partial_{\eta_2}U+\mu_2\partial^2_{\eta_2}U.
\label{eqn: BKE_(0,1)}
\end{eqnarray}
Separation of variables $U(\eta_1,\eta_2)=X_1(\eta_1)X_2(\eta_2)$ reduces
Eq. (\ref{eqn: BKE_(0,1)})  to  an eigenvalue problem 
\begin{eqnarray}
\lambda X_1 &&= \Big(\frac{\mu_1+\mu_2}{2}-\frac{\mu_1 \beta_1}{2}\Big)\eta_1X''_1+\mu_1
\beta_1\eta_1X'_1,\label{SepVar_x1} \\
-\lambda X_2 &&=\mu_2X''_2-\mu_2\eta_2X'_2, \label{SepVar_X2}
\end{eqnarray}
where $\lambda$ is an eigenvalue. The general solution to 
(\ref{SepVar_X2}) is
\begin{equation} 
X_2(\eta_2)=C_{1}H_{\frac{\lambda}{\mu_2}}\left(\frac{\eta_{2}}{\sqrt{2}}\right)
+C_2 {}_{1}F_1\left(-\frac{\lambda}{2\mu_2};
\frac{1}{2};\frac{\eta_2^2}{2}\right), 
\end{equation} where
$C_1$ and $C_2$ are constants, $H_{n}(z)$ is the Hermite
polynomial, and ${}_{1}F_{1}(a;b;z) $ is the confluent hypergeometric function
of the first kind.  The matching condition to the  plateau fixation probability
$U(\eta_1 \rightarrow \infty,\eta_2)=P$ enforces $X_{2}(\eta_2)$
to be  a constant. This is possible only when $C_{2}=0$ and $\lambda
=0 $ so  $X_{2}(\eta_2) = C_1.$ Because zero is the only eigenvalue
consistent with the matching condition, (\ref{SepVar_x1}) reduces to  
\begin{equation}
0 = \Big(\frac{\mu_1+\mu_2}{2}-\frac{\mu_1 \beta_1}{2}\Big)\eta_1X''_1+\mu_1
\beta_1\eta_1X'_1,
\label{SepVarZeroLambda_x1} 
\end{equation}
whose general solution is 
\begin{equation}
X_1(\eta_1)=B_1-B_{2} \Big(\frac{\mu_1+\mu_2}{2\mu_1
\beta_1}-\frac{1}{2}\Big)e^{-\eta_1
\mu_1 \beta_1/\left(\mu_1+\mu_2-\mu_1 \beta_1\right)},
\end{equation}
where $B_{1} $
and $B_{2}$ are constants.
 By imposing the boundary condition
$U(0,\eta_2)=0$   and the matching condition 
$U(\eta_1\rightarrow \infty,\eta_2)=P,$ it follows that the fixation
probability in the original coordinates valid in the vicinity of the fixed
point $(0,1)$ is  
\begin{equation}
u(\boldsymbol c) = P +P  e^{-N c_1 [2\beta_1/(2-\beta_1)]}.
\label{eqn: AsympFP(0,1)}
\end{equation}
A similar argument can be applied to the asymptotic solution near the fixed
point $(1,0).$ In this case, the stretched coordinates are $\eta_1=(c_1-1)\sqrt{N}$
and $ \eta_2=c_2N,$ with the fixation probability in the new coordinates
given by $U(\eta_1,\eta_2)=u(1+\eta_1/\sqrt{N},\eta_2/N).$ Eq. (\ref{eqn:
BKE_2d}) in the new coordinates,
with terms of
$\mathcal{O}(1/\sqrt{N})$ neglected, reads
\begin{eqnarray}
0=\ &&-\mu_1\eta_1\partial_{\eta_1}U+\mu_1\partial^2_{\eta_1}U +\mu_2 \beta_2\eta_2\partial_{\eta_2}U\nonumber\\
&&+ \Big(\frac{\mu_1+\mu_2}{2}-\frac{\mu_2 \beta_2}{2}\Big)\eta_2\partial^2_{\eta_2}U,
\label{eqn: BKE_(1,0)}
\end{eqnarray}
which is equivalent to  (\ref{eqn: BKE_(0,1)})
 with indices 1 and 2 interchanged. Following the method of separation
of variables as above and imposing the boundary condition
$U(\eta_1,0) =1$ as well as the matching 
condition $U(\eta_1,\eta_2\rightarrow \infty)=P,$ we arrive at the fixation
probability valid in the vicinity of the
fixed point $(1,0)$ 
\begin{equation}
u(\boldsymbol c) = P+(1-P) e^{-N c_2 [2\beta_2/(2-\beta_2)]}.
\label{eqn: AsympFP(1,0)}
\end{equation}
        
        Now consider  the asymptotic solutions away from the saddle
fixed points but still in the boundary layers. In the boundary layer adjacent
to the absorbing boundary $c_{1}=0$ but away from the saddle fixed point
$(0,1)$, we introduce the stretched coordinate $\eta_1 = c_1N$ and $\eta_2
= c_{2}$. Upon neglecting the contributions of $\mathcal{O}(1/N)$ and rewriting
the fixation probability in the new coordinate as
$U(\eta_1,\eta_2)=u(\eta_1/N,\eta_2)$,  Eq. (\ref{eqn:
BKE_2d}) becomes
\begin{eqnarray}
0=\ &&2\mu_2\eta_2(1-\eta_2)\partial_{\eta_2}U+2[\mu_1-(\mu_1-\mu_1 \beta_1)\eta_2]\eta_1\partial_{\eta_1}U\nonumber\\
&&+[\mu_1-(\mu_1+\mu_1
\beta_1)\eta_2]\eta_1\partial_{\eta_1}^2
U. \label{eqn: BKEStretched_BL}
\end{eqnarray}
 We can turn (\ref{eqn: BKEStretched_BL}) into a separable PDE and solve
the associated eigenvalue problem by transforming to the new coordinates
$x_{1}=\eta_1/\Phi_2(\eta_2), $ and $x_2=\eta_2$. Substituting the
coordinate transformation $V(x_1,x_2)=U( x_1\Phi_{2}(x_2),x_2)$  into (\ref{eqn:
BKEStretched_BL}), we find $V(x_1,x_2)$ satisfies
a separable PDE
\begin{eqnarray}
0 = \ &&x_1\partial^2_{x_1}V + x_1\partial_{x_1}V \nonumber\\
&&+ 2\frac{\mu_2}{\mu_1}\Big[\frac{x_2(1-x_2)}{1
- (1+\beta_1)x_2}\Phi_2(x_2)\Big] \partial_{x_2}V,\label{eqn: sepV}
\end{eqnarray}
with $\Phi_2(x)$ obeys the first order differential equation given in
Eq (\ref{eqn: main_phi2}).
Separation of variables $V(x_1,x_2) = V_1(x_1)V_2(x_2)$ turns (\ref{eqn:
sepV}) into an eigenvalue problem 
\begin{eqnarray}
\lambda V_1 &&=x_{1}V_{1}''+x_{1}V'_{1},
\label{SepVar_v1} \\
-\lambda V_2 &&= 2\frac{\mu_2}{\mu_1}\Big[\frac{x_2(1-x_2)}{1
- (1+\beta_1)x_2}\Phi_2(x_2)\Big] V'_{2}, \label{SepVar_v2}
\end{eqnarray}
with $\lambda $ the eigenvalue. Again, matching to the fixation probability
at the plateau $U(\eta_1\rightarrow\infty,\eta_2)=P$ enforces $V_2(x_2)$
to be  constant which is possible only if $\lambda=0.$  The general
solution to (\ref{SepVar_v1}) with $\lambda=0$ is 
\begin{equation}
V_1(x_1)=D_1-D_{2}e^{-x_{1}}.
\end{equation}
Upon imposing the boundary condition $U(0,\eta_2)=0$ as well as the matching
condition $U(\eta_1\rightarrow\infty,\eta_2)=P,$ we obtain the fixation
probability in the original coordinate valid within the boundary layer
adjacent to the absorbing boundary $c_{1}=0,$ namely
\begin{equation}
u(\boldsymbol c) = P -P  e^{-N c_1 /\Phi_2(c_2)}.\label{eqn: AsymptoticC2Fixed}
\end{equation}
Upon matching   (\ref{eqn: AsymptoticC2Fixed})  to the asymptotic solution
in the vicinity of the saddle fixed point $(1,0),$ (\ref{eqn: AsympFP(1,0)}),
we find a  first order differential equation governing $\Phi_2,$
given by Eq. (\ref{eqn: main_phi2}), with the matching condition $\lim_{x\rightarrow
1} \Phi_2(x) =\frac{2-\beta_1}{2\beta_1}$.
 
A similar argument with index 1 and 2 interchanged determines the asymptotic
solution within the boundary layer adjacent
to the absorbing boundary $c_{2}=0.$ We find that the fixation probability
in this region is given by 
\begin{equation}
u(\boldsymbol c) = P +(1-P)  e^{-N c_2 /\Phi_1(c_1)},\label{eqn: AsymptoticC1Fixed}
\end{equation}
where $\Phi_1(x)$ obeys Eq. (\ref{eqn: main_phi1}) subject to the matching
condition $\lim_{x \rightarrow 1} \Phi_1(x) = \frac{2-\beta_2}{2\beta_2}$.
Therefore, the global solution with  smooth crossovers from the plateau
\textit{P }to the the boundary layer behavior of (\ref{eqn: AsymptoticC2Fixed})
and (\ref{eqn: AsymptoticC1Fixed})
is given by Eq. (\ref{eqn: FixProb_Mutualism_c1c2}).

\section{Fixation from a quasi-stationary distribution}
\label{sec:plateau_QSD}
The analysis in this section follows  the general discussion  on the  high
dimensional exit
problem
  by Grasman and Herwaarden \cite{Grasman:1999kx}. We first argue that, near
the absorbing
boundaries, the quasi-stationary distribution (QSD) $p_{st}(\boldsymbol
c)$ is peaked at the saddle fixed point $(0,1)$ and $(1,0).$  To see this,
consider the (stationary) Fokker-Planck equation 
\begin{equation}
0=\sum_{i=1}^2\Big(-\partial_{c_i}v_ip_{st}
+ \frac{1}{2N}\partial_{c_i}^2D_ip_{st}
\Big),\label{eqn: Stationary Fokker-Planck}
\end{equation}
where $v_i$ and $D_i$ are given by Eqs. (\ref{eqn: drift1})-(\ref{eqn: diff2}).
When the problem can be regarded as a rare event escape from
a metastable state, the  asymptotic solution to (\ref{eqn: Stationary
Fokker-Planck})  is solved by the WKB ansatz \cite{Dykman:1994pz,Assaf:2010xp,Graham:1985rf,Grasman:1999kx}
\begin{equation}
p_{st}(\boldsymbol c) = w(\boldsymbol c)e^{-N\Psi(\boldsymbol c)}.
\label{eqn: WKB}
\end{equation} 
Substituting  (\ref{eqn: WKB}) into (\ref{eqn: Stationary Fokker-Planck})
and collecting the leading order terms in $N$ leads to an eikonal equation,
 
\begin{equation}
0=\sum_{i=1}^2\Big(v_{i}(\partial_{c_i}\Psi)
+ \frac{1}{2}D_{i}(\partial_{c_i} 
\Psi)^{2}\Big).
\label{eqn: eikonal}
\end{equation}
Collecting terms of $\mathcal{O}(1)$ results in 
\begin{eqnarray}
0=\ &&\sum_{i=1}^2\Big (D_{i }(\partial_{c_i}\Psi)+
v_i\Big)\partial_{c_i}w\nonumber\\
&&+\sum_{i=1}^2\Big(\frac{1}{2}D_i\partial^2_{c_i}\Psi+
(\partial_{c_i}D_i)(\partial_{c_i}\Psi)\Big)w.
\label{eqn: eqnforW}
\end{eqnarray}
In the neighborhood of the absorbing boundary  $c_1=0$, we expand $\Psi$
around $c_{1}=0$ as
\begin{equation}
\Psi(\boldsymbol c)=\Psi^{(0)}_2(c_2)+\Psi^{(1)}_2(c_2)c_1+\frac{1}{2}\Psi^{(2)}_2(c_2)c_1^2+\cdots
,\label{eqn: PsiExpand_absbound}
\end{equation}
where the subscript 2 of $\Psi$ denotes the expansion around the fixation
of species 2 and the superscript labels the order of expansion. Upon  substituting
the expansion (\ref{eqn: PsiExpand_absbound}) into
(\ref{eqn: eikonal}) and collecting terms of $\mathcal{O}(c^0_1),$ we
find 
$\Psi'^{(0)}_{2}(c_2)=2(c_2-1)/(c_2+1).$ Therefore, in the limit  $c_1
\rightarrow 0,$ $\Psi(\boldsymbol c)$ is minimal at $c_2=1,$ implying
that $p_{st}$ is peaked in the neighborhood of the fixed point $(0,1)$
 provided $N\gg 1.$      

For the behavior of $w(\boldsymbol c)$ near $(0,1),$  it turns out
the singular behavior of $w$ when $c_{1}\rightarrow0 $ scales as $w\sim1/c_1.$
We refer to the discussion in Ref. \cite{Grasman:1999kx} for the related
problem of extinction probability in the predator-prey model. The singular
behavior suggests the QSD is concentrated in the neighborhood of the saddle
fixed point. 

To extract the quantitative behavior of $p_{st}$ near the saddle fixed
point $(0,1)$,
we Taylor expand $\Psi$ around $(0,1)$ 
\begin{equation}
\Psi(\boldsymbol c)=\bar\Psi^{(0)}_2+\bar\Psi^{(1)}_2c_1+\bar\Psi^{(2)}_2(c_2-1)+\frac{1}{2}\bar\Psi^{(3)}_2(c_2-1)^2+\cdots
,\label{eqn: PsiExpand(0,1)}
\end{equation}
where we denote  the $i^{\textrm{th}}$ expansion coefficient around the
saddle fixed point of species 2 by $\bar\Psi^{(i)}_2$ . Upon substituting
the expansion (\ref{eqn: PsiExpand(0,1)}) into
(\ref{eqn: eikonal}) and (\ref{eqn: eqnforW}) we get $\bar\Psi^{(1)}_2=-2\beta_1/(2-\beta_1)$,
$\bar\Psi^{(2)}_2=0,$
and $\bar\Psi^{(3)}_2=1.$
Therefore, in the neighborhood of the fixed point $(0,1),$ the QSD takes
the form 
\begin{eqnarray}
p_{st}(\boldsymbol c)\approx &&\frac{\bar w^{(0)}_2\exp(-N\bar\Psi^{(0)}_2)}{c_1}\nonumber\\
&&\times\exp\left[N\left(\frac{2\beta_1}{2-\beta_1}c_1-\frac{(c_{2}-1)^{2}}{2}\right)\right].
\label{eqn: WKB_(0,1)}
\end{eqnarray}
Similar arguments lead to the behavior of the QSD in the neighborhood of
the fixed point $(1,0), $ which reads
\begin{eqnarray}
p_{st}(\boldsymbol c)\approx &&\frac{\bar w^{(0)}_1\exp(-N\bar\Psi^{(0)}_1)}{c_2}\nonumber\\
&&\times\exp\left[N\left(\frac{2\beta_2}{2-\beta_2}c_2-\frac{(c_{1}-1)^{2}}{2}\right)\right].
\label{eqn: WKB_(1,0)}
\end{eqnarray}

We now relate  the behavior of the QSD near the absorbing boundaries to
the plateau fixation probability $P$ in the bulk region by  employing the
 identity resulting
from the divergence theorem:  
\begin{eqnarray}
&&\int_{\Omega}(p\hat Lu-u\hat Mp)dc_1dc_2  \nonumber\\
&&=
\int_{\partial \Omega}\bigg(\sum_{i=1}^2 \frac{1}{2N}\Big[n_iD_i(p\partial_{c_i}u-u\partial_{c_i}p)-n_i(\partial_{c_i}D_i)pu\Big]\nonumber\\
&&\hspace{1.2cm}+\sum_{i=1}^2n_iv_ipu\bigg)dS,
\label{eqn: divergence}
\end{eqnarray}
where $\hat L$ is the backward-Kolmogorov operator, $u$ is the solution
to the backward-Kolmogorov equation, $\hat M$ is the forward-Kolmogorov
(Fokker-Planck)
operator, $p$ is the solution
to the forward-Kolmogorov equation, $\Omega$ is
the domain of interest, and $n_i$ is the $i^{th}$ components of the
normal vector at the boundary $\partial\Omega. $ 
In the long-time limit when the QSD  has already
developed, the volume integral (left hand side) of (\ref{eqn: divergence})
vanishes since $\hat Lu=0$ and $\hat Mp_{st}=0$.
To evaluate the surface integral   in  (\ref{eqn: divergence}) and avoid
the singularity
of $p_{st}$ on each absorbing boundary, we consider the domain $\Omega
=\{\boldsymbol c \hspace{1mm}|\hspace{1mm}c_1>\varepsilon,
c_2>\varepsilon\}$ and evaluate  (\ref{eqn: divergence}) in   the limit
$\varepsilon \rightarrow 0$.
In this domain, (\ref{eqn: divergence}) becomes
\begin{eqnarray}
0 =
&&\int_\varepsilon^\infty dc_{2}\bigg( \frac{1}{2N} \big[D_1(p_{st}\partial_{c_1}u-u\partial_{c_1}p_{st})-(\partial_{c_1}D_1)p_{st}u\big]\nonumber\\
&&\hspace{1.7cm} + \ v_1p_{st}u\bigg)_{c_1=\varepsilon}\nonumber\\
&&+\int_\varepsilon^\infty dc_{1}\bigg(\frac{1}{2N} \big[D_2(p_{st}\partial_{c_2}u-u\partial_{c_2}p_{st})-(\partial_{c_2}D_2)p_{st}u\big]\nonumber\\
&&\hspace{1.8cm}+ \ v_2p_{st}u\bigg)_{c_2=\varepsilon}.
\label{eqn: divergence_QSD}
\end{eqnarray}
(\ref{eqn: divergence_QSD}) relates the plateau fixation probability $P$
contained in $u$ by   (\ref{eqn: AsympFP(0,1)}) and
(\ref{eqn: AsympFP(1,0)}) to the boundary behavior of $p_{st}$.
Substituting the asymptotic solutions of the QSD
given by (\ref{eqn: WKB_(0,1)}) and (\ref{eqn: WKB_(1,0)}), the asymptotic
solutions of $u$ given by (\ref{eqn: AsympFP(0,1)}) and
(\ref{eqn: AsympFP(1,0)}),
and the deterministic drifts as well as diffusion coefficients given by Eqs.
(\ref{eqn:
drift1})-(\ref{eqn: diff2}) into (\ref{eqn:
divergence_QSD}), we obtain after taking the limits $\varepsilon \rightarrow0$
and $N \gg 1$
\begin{eqnarray}
0 = \ &&\frac{2\beta_1}{2-\beta_1} \bar w^{(0)}_2\exp(-N\bar\Psi^{(0)}_2)\nonumber\\
&&\hspace{2mm}\times\int_0^\infty dc_{2}\bigg\{\left[-\frac{\mu_1P }{2}-\frac{\mu_1(1-\beta_1)P}{2}c_2
\right]\nonumber\\
&&\hspace{2.8cm}\times\exp\left[-\frac{N(c_2-1)^2}{2}\right]\bigg\}\nonumber\\
&& + \ \frac{2\beta_2}{2-\beta_2}\bar w^{(0)}_1\exp(-N\bar\Psi^{(0)}_1)\nonumber\\
&& \hspace{4mm}\times\int_0^\infty dc_{1}\bigg\{\bigg[\mu_2\left(1-\frac{2-\beta_2}{2\beta_2}-\frac{P}{2}\right)\nonumber\\
&&\hspace{2.4cm}+\mu_2(1-\beta_2)\left(1+\frac{2-\beta_2}{2\beta_2}-\frac{P}{2}\right)c_1\bigg]\nonumber\\
&& \hspace{2.9cm} \times \exp\left[-\frac{N(c_1-1)^2}{2}\right]\bigg\}.
\label{eqn: NeedLaplace}
\end{eqnarray} 
The integrals can be evaluated by the standard method
of Laplace integration when $N \gg 1$. The result reads
\begin{eqnarray}
0 = \ && \frac{2\beta_1}{2-\beta_1}\bar w^{(0)}_2\exp(-N\bar\Psi^{(0)}_2)\sqrt{\frac{\pi}{N}}\nonumber\\
&&\hspace{0.7cm}\times\left[-\frac{\mu_1P}{2}-\frac{\mu_1(1-\beta_1)P}{2}\right]\nonumber\\
&& +\frac{2\beta_2}{2-\beta_2}\bar w^{(0)}_1\exp(-N\bar\Psi^{(0)}_1)\sqrt{\frac{\pi}{N}}\nonumber\\
&&\hspace{0.7cm}\times\bigg[\mu_2\left(1-\frac{2-\beta_2}{2\beta_2}-\frac{P}{2}\right)\nonumber\\
&&\hspace{1.05cm}+\mu_2(1-\beta_2)\left(1+\frac{2-\beta_2}{2\beta_2}-\frac{P}{2}\right)\bigg].
\label{eqn: AfterLaplace}
\end{eqnarray}
\newline
Upon rewriting $\beta_1 = \mu_1-\lambda_{12}N $ , and $\beta_2 = \mu_2-\lambda_{21}N$
 and keeping only the leading order term in $1/N$, we obtain the plateau
fixation probability
\begin{equation}
P \approx \frac{\lambda_{21}\bar w^{(0)}_1e^{-N\bar\Psi^{(0)}_1}}{\lambda_{21}\bar
w^{(0)}_1e^{-N\bar\Psi^{(0)}_1}+\lambda_{12}\bar w^{(0)}_2e^{-N\bar\Psi^{(0)}_2}}.\label{eqn:Appndx_PlateauFP}
\end{equation}
\newline 
Recall that $\bar w_1^{(0)}e^{-N\bar\Psi_1^{(0)}}$ and  $\bar w_2^{(0)}e^{-N\bar\Psi_2^{(0)}}$
 are    $p_{st}(\boldsymbol c)$ evaluated at  $(1,0^{+})$ and $(0^{+},1).$
Consequently, (\ref{eqn:Appndx_PlateauFP}) is the ratio of the flux into
 (1,0) to the total flux into 
(1,0) and (0,1).  In the limit   $N \gg 1,$  (\ref{eqn: WKB_(0,1)})
and (\ref{eqn: WKB_(1,0)}) imply that, on each absorbing boundary,  the 
QSD peaks up  at the saddle fixed point while the width around the peak becomes
vanishingly narrow;
accordingly, the flux into the saddle fixed point well approximates the
flux into the corresponding absorbing boundary. Hence, (\ref{eqn:Appndx_PlateauFP})
describes the ratio of flux into the absorbing boundary at $f=1$
to the total
flux into both the absorbing boundaries at $f=0$ and $f=1$.    

 Note that (\ref{eqn:Appndx_PlateauFP}) can be rewritten in the form similar
to Eq. (\ref{eqn:plateau_height}) as
\begin{equation}
P \approx \frac{1}{1+e^{-N\Delta S_0 + \Delta S_1} },
\end{equation}
where $\Delta S_0 \equiv \bar\Psi_2^{(0)} - \bar\Psi_1^{(0)} $
and $\Delta S_1 \equiv \ln (\lambda_{12}\bar w_2^{(0)})-\ln (\lambda_{21}\bar
w_1^{(0)}).$ Since   $\bar w_i^{(0)}$ and $\bar\Psi_i^{(0)}$ are independent
of $N,$ we can vary $N$ while fixing  $s_{o}$, $\beta_1$ and $\beta_2$ to
infer  $\Delta S_0$ and $\Delta S_1$ by fitting the plateau fixation probability
$P$
to simulations.
In principle, the  exact values of $\Delta S_0$ and $\Delta S_1$
 may be obtained numerically by simultaneously solving    $\bar w_i^{(0)}$
and $\bar\Psi_i^{(0)}$    from (\ref{eqn:
eikonal}) and (\ref{eqn: eqnforW}) \cite{Kamenev:2008xu,Assaf:2010eb,Dykman:1994pz,Elgart:2004ya},
but these are beyond the scope of this work.       


\begin{thebibliography}{61}
\expandafter\ifx\csname natexlab\endcsname\relax\def\natexlab#1{#1}\fi
\expandafter\ifx\csname bibnamefont\endcsname\relax
  \def\bibnamefont#1{#1}\fi
\expandafter\ifx\csname bibfnamefont\endcsname\relax
  \def\bibfnamefont#1{#1}\fi
\expandafter\ifx\csname citenamefont\endcsname\relax
  \def\citenamefont#1{#1}\fi
\expandafter\ifx\csname url\endcsname\relax
  \def\url#1{\texttt{#1}}\fi
\expandafter\ifx\csname urlprefix\endcsname\relax\def\urlprefix{URL }\fi
\providecommand{\bibinfo}[2]{#2}
\providecommand{\eprint}[2][]{\url{#2}}

\bibitem[{\citenamefont{Elena and Lenski}(2003)}]{Elena:2003sf}
\bibinfo{author}{\bibfnamefont{S.~F.} \bibnamefont{Elena}} \bibnamefont{and}
  \bibinfo{author}{\bibfnamefont{R.~E.} \bibnamefont{Lenski}},
  \bibinfo{journal}{Nature Reviews Genetics} \textbf{\bibinfo{volume}{4}},
  \bibinfo{pages}{457} (\bibinfo{year}{2003}).

\bibitem[{\citenamefont{Desai}(2013)}]{Desai:2013jp}
\bibinfo{author}{\bibfnamefont{M.~M.} \bibnamefont{Desai}},
  \bibinfo{journal}{Journal of Statistical Mechanics: Theory and Experiment}
  \textbf{\bibinfo{volume}{2013}}, \bibinfo{pages}{P01003}
  (\bibinfo{year}{2013}).

\bibitem[{\citenamefont{Dai et~al.}(2012)\citenamefont{Dai, Vorselen, Korolev,
  and Gore}}]{Dai:2012wd}
\bibinfo{author}{\bibfnamefont{L.}~\bibnamefont{Dai}},
  \bibinfo{author}{\bibfnamefont{D.}~\bibnamefont{Vorselen}},
  \bibinfo{author}{\bibfnamefont{K.~S.} \bibnamefont{Korolev}},
  \bibnamefont{and} \bibinfo{author}{\bibfnamefont{J.}~\bibnamefont{Gore}},
  \bibinfo{journal}{Science} \textbf{\bibinfo{volume}{336}},
  \bibinfo{pages}{1175} (\bibinfo{year}{2012}).

\bibitem[{\citenamefont{Sanchez and Gore}(2013)}]{Sanchez:2013eu}
\bibinfo{author}{\bibfnamefont{A.}~\bibnamefont{Sanchez}} \bibnamefont{and}
  \bibinfo{author}{\bibfnamefont{J.}~\bibnamefont{Gore}},
  \bibinfo{journal}{PLoS biology} \textbf{\bibinfo{volume}{11}},
  \bibinfo{pages}{e1001547} (\bibinfo{year}{2013}).

\bibitem[{\citenamefont{Griffin et~al.}(2004)\citenamefont{Griffin, West, and
  Buckling}}]{Griffin:2004qe}
\bibinfo{author}{\bibfnamefont{A.~S.} \bibnamefont{Griffin}},
  \bibinfo{author}{\bibfnamefont{S.~A.} \bibnamefont{West}}, \bibnamefont{and}
  \bibinfo{author}{\bibfnamefont{A.}~\bibnamefont{Buckling}},
  \bibinfo{journal}{Nature} \textbf{\bibinfo{volume}{430}},
  \bibinfo{pages}{1024} (\bibinfo{year}{2004}).

\bibitem[{\citenamefont{Nowak}(2006)}]{Nowak:2006uq}
\bibinfo{author}{\bibfnamefont{M.~A.} \bibnamefont{Nowak}},
  \emph{\bibinfo{title}{Evolutionary Dynamics: Exploring the Equations of
  Life}} (\bibinfo{publisher}{Harvard University Press}, \bibinfo{year}{2006}).

\bibitem[{\citenamefont{Frey}(2010)}]{Frey:2010fk}
\bibinfo{author}{\bibfnamefont{E.}~\bibnamefont{Frey}},
  \bibinfo{journal}{Physica A: Statistical Mechanics and its Applications}
  \textbf{\bibinfo{volume}{389}}, \bibinfo{pages}{4265} (\bibinfo{year}{2010}).

\bibitem[{\citenamefont{Blythe and McKane}(2007)}]{Blythe:2007nr}
\bibinfo{author}{\bibfnamefont{R.~A.} \bibnamefont{Blythe}} \bibnamefont{and}
  \bibinfo{author}{\bibfnamefont{A.~J.} \bibnamefont{McKane}},
  \bibinfo{journal}{Journal of Statistical Mechanics: Theory and Experiment}
  \textbf{\bibinfo{volume}{2007}}, \bibinfo{pages}{P07018}
  (\bibinfo{year}{2007}).

\bibitem[{\citenamefont{Ewens}(2004)}]{Ewens:2004kx}
\bibinfo{author}{\bibfnamefont{W.~J.} \bibnamefont{Ewens}},
  \emph{\bibinfo{title}{Mathematical Population Genetics 1: I. Theoretical
  Introduction}} (\bibinfo{publisher}{Springer}, \bibinfo{year}{2004}), ISBN
  \bibinfo{isbn}{0387201912}.

\bibitem[{\citenamefont{Gillespie}(2010)}]{Gillespie:2010uq}
\bibinfo{author}{\bibfnamefont{J.~H.} \bibnamefont{Gillespie}},
  \emph{\bibinfo{title}{Population Genetics: A Concise Guide}}
  (\bibinfo{publisher}{JHU Press}, \bibinfo{year}{2010}), ISBN
  \bibinfo{isbn}{1421401703}.

\bibitem[{\citenamefont{Hartl et~al.}(1997)\citenamefont{Hartl, Clark
  et~al.}}]{Hartl:1997hb}
\bibinfo{author}{\bibfnamefont{D.~L.} \bibnamefont{Hartl}},
  \bibinfo{author}{\bibfnamefont{A.~G.} \bibnamefont{Clark}},
  \bibnamefont{et~al.}, \emph{\bibinfo{title}{Principles of Population
  Genetics}}, vol. \bibinfo{volume}{116} (\bibinfo{publisher}{Sinauer
  associates Sunderland}, \bibinfo{year}{1997}).

\bibitem[{\citenamefont{Lambert et~al.}(2014)\citenamefont{Lambert, Vyawahare,
  and Austin}}]{Lambert:2014aa}
\bibinfo{author}{\bibfnamefont{G.}~\bibnamefont{Lambert}},
  \bibinfo{author}{\bibfnamefont{S.}~\bibnamefont{Vyawahare}},
  \bibnamefont{and} \bibinfo{author}{\bibfnamefont{R.~H.}
  \bibnamefont{Austin}}, \bibinfo{journal}{Interface focus}
  \textbf{\bibinfo{volume}{4}}, \bibinfo{pages}{20140029}
  (\bibinfo{year}{2014}).

\bibitem[{\citenamefont{Smith}(1982)}]{Smith:1982it}
\bibinfo{author}{\bibfnamefont{J.~M.} \bibnamefont{Smith}},
  \emph{\bibinfo{title}{Evolution and the Theory of Games}}
  (\bibinfo{publisher}{Cambridge university press}, \bibinfo{year}{1982}).

\bibitem[{\citenamefont{Korolev and Nelson}(2011)}]{Korolev:2011vn}
\bibinfo{author}{\bibfnamefont{K.~S.} \bibnamefont{Korolev}} \bibnamefont{and}
  \bibinfo{author}{\bibfnamefont{D.~R.} \bibnamefont{Nelson}},
  \bibinfo{journal}{Physical Review Letters} \textbf{\bibinfo{volume}{107}},
  \bibinfo{pages}{088103} (\bibinfo{year}{2011}).

\bibitem[{\citenamefont{M{\"u}ller et~al.}(2014)\citenamefont{M{\"u}ller,
  Neugeboren, Nelson, and Murray}}]{Muller:2014wo}
\bibinfo{author}{\bibfnamefont{M.~J.} \bibnamefont{M{\"u}ller}},
  \bibinfo{author}{\bibfnamefont{B.~I.} \bibnamefont{Neugeboren}},
  \bibinfo{author}{\bibfnamefont{D.~R.} \bibnamefont{Nelson}},
  \bibnamefont{and} \bibinfo{author}{\bibfnamefont{A.~W.}
  \bibnamefont{Murray}}, \bibinfo{journal}{Proceedings of the National Academy
  of Sciences} \textbf{\bibinfo{volume}{111}}, \bibinfo{pages}{1037}
  (\bibinfo{year}{2014}).

\bibitem[{\citenamefont{Kimura}(1984)}]{Kimura:1984ph}
\bibinfo{author}{\bibfnamefont{M.}~\bibnamefont{Kimura}},
  \emph{\bibinfo{title}{The Neutral Theory of Molecular Evolution}}
  (\bibinfo{publisher}{Cambridge University Press}, \bibinfo{year}{1984}).

\bibitem[{\citenamefont{Korolev et~al.}(2010)\citenamefont{Korolev, Avlund,
  Hallatschek, and Nelson}}]{Korolev:2010fk}
\bibinfo{author}{\bibfnamefont{K.}~\bibnamefont{Korolev}},
  \bibinfo{author}{\bibfnamefont{M.}~\bibnamefont{Avlund}},
  \bibinfo{author}{\bibfnamefont{O.}~\bibnamefont{Hallatschek}},
  \bibnamefont{and} \bibinfo{author}{\bibfnamefont{D.~R.}
  \bibnamefont{Nelson}}, \bibinfo{journal}{Reviews of Modern Physics}
  \textbf{\bibinfo{volume}{82}}, \bibinfo{pages}{1691} (\bibinfo{year}{2010}).

\bibitem[{\citenamefont{Kimura}(1962)}]{Kimura:1962ez}
\bibinfo{author}{\bibfnamefont{M.}~\bibnamefont{Kimura}},
  \bibinfo{journal}{Genetics} \textbf{\bibinfo{volume}{47}},
  \bibinfo{pages}{713} (\bibinfo{year}{1962}).

\bibitem[{\citenamefont{Moran et~al.}(1962)}]{Moran:1962pi}
\bibinfo{author}{\bibfnamefont{P.~A.~P.} \bibnamefont{Moran}}
  \bibnamefont{et~al.}, \bibinfo{journal}{The Statistical Processes of
  Evolutionary Theory.}  (\bibinfo{year}{1962}).

\bibitem[{\citenamefont{Cremer et~al.}(2009)\citenamefont{Cremer, Reichenbach,
  and Frey}}]{Cremer:2009sh}
\bibinfo{author}{\bibfnamefont{J.}~\bibnamefont{Cremer}},
  \bibinfo{author}{\bibfnamefont{T.}~\bibnamefont{Reichenbach}},
  \bibnamefont{and} \bibinfo{author}{\bibfnamefont{E.}~\bibnamefont{Frey}},
  \bibinfo{journal}{New Journal of Physics} \textbf{\bibinfo{volume}{11}},
  \bibinfo{pages}{093029} (\bibinfo{year}{2009}).

\bibitem[{\citenamefont{Mobilia and Assaf}(2010)}]{Mobilia:2010fk}
\bibinfo{author}{\bibfnamefont{M.}~\bibnamefont{Mobilia}} \bibnamefont{and}
  \bibinfo{author}{\bibfnamefont{M.}~\bibnamefont{Assaf}},
  \bibinfo{journal}{EPL (Europhysics Letters)} \textbf{\bibinfo{volume}{91}},
  \bibinfo{pages}{10002} (\bibinfo{year}{2010}).

\bibitem[{\citenamefont{Assaf and Mobilia}(2010)}]{Assaf:2010eb}
\bibinfo{author}{\bibfnamefont{M.}~\bibnamefont{Assaf}} \bibnamefont{and}
  \bibinfo{author}{\bibfnamefont{M.}~\bibnamefont{Mobilia}},
  \bibinfo{journal}{Journal of Statistical Mechanics: Theory and Experiment}
  \textbf{\bibinfo{volume}{2010}}, \bibinfo{pages}{P09009}
  (\bibinfo{year}{2010}).

\bibitem[{\citenamefont{Lassig}(2002)}]{Lassig:2002aa}
\bibinfo{author}{\bibfnamefont{M.}~\bibnamefont{Lassig}},
  \bibinfo{journal}{arXiv preprint cond-mat/0206093}  (\bibinfo{year}{2002}).

\bibitem[{\citenamefont{Lavrentovich and Nelson}(2014)}]{Lavrentovich:2014oj}
\bibinfo{author}{\bibfnamefont{M.~O.} \bibnamefont{Lavrentovich}}
  \bibnamefont{and} \bibinfo{author}{\bibfnamefont{D.~R.}
  \bibnamefont{Nelson}}, \bibinfo{journal}{Physical Review Letters}
  \textbf{\bibinfo{volume}{112}}, \bibinfo{pages}{138102}
  (\bibinfo{year}{2014}).

\bibitem[{\citenamefont{Lavrentovich et~al.}(2013)\citenamefont{Lavrentovich,
  Korolev, and Nelson}}]{Lavrentovich:2013ys}
\bibinfo{author}{\bibfnamefont{M.~O.} \bibnamefont{Lavrentovich}},
  \bibinfo{author}{\bibfnamefont{K.~S.} \bibnamefont{Korolev}},
  \bibnamefont{and} \bibinfo{author}{\bibfnamefont{D.~R.}
  \bibnamefont{Nelson}}, \bibinfo{journal}{Physical Review E}
  \textbf{\bibinfo{volume}{87}}, \bibinfo{pages}{012103}
  (\bibinfo{year}{2013}).

\bibitem[{\citenamefont{Hutchinson}(1961)}]{Hutchinson:1961bf}
\bibinfo{author}{\bibfnamefont{G.~E.} \bibnamefont{Hutchinson}},
  \bibinfo{journal}{American Naturalist} pp. \bibinfo{pages}{137--145}
  (\bibinfo{year}{1961}).

\bibitem[{\citenamefont{T{\'e}l et~al.}(2005)\citenamefont{T{\'e}l, de~Moura,
  Grebogi, and K{\'a}rolyi}}]{Tel:2005kk}
\bibinfo{author}{\bibfnamefont{T.}~\bibnamefont{T{\'e}l}},
  \bibinfo{author}{\bibfnamefont{A.}~\bibnamefont{de~Moura}},
  \bibinfo{author}{\bibfnamefont{C.}~\bibnamefont{Grebogi}}, \bibnamefont{and}
  \bibinfo{author}{\bibfnamefont{G.}~\bibnamefont{K{\'a}rolyi}},
  \bibinfo{journal}{Physics Reports} \textbf{\bibinfo{volume}{413}},
  \bibinfo{pages}{91} (\bibinfo{year}{2005}).

\bibitem[{\citenamefont{Perlekar et~al.}(2010)\citenamefont{Perlekar, Benzi,
  Nelson, and Toschi}}]{Perlekar:2010lk}
\bibinfo{author}{\bibfnamefont{P.}~\bibnamefont{Perlekar}},
  \bibinfo{author}{\bibfnamefont{R.}~\bibnamefont{Benzi}},
  \bibinfo{author}{\bibfnamefont{D.~R.} \bibnamefont{Nelson}},
  \bibnamefont{and} \bibinfo{author}{\bibfnamefont{F.}~\bibnamefont{Toschi}},
  \bibinfo{journal}{Physical Review Letters} \textbf{\bibinfo{volume}{105}},
  \bibinfo{pages}{144501} (\bibinfo{year}{2010}).

\bibitem[{\citenamefont{Pigolotti et~al.}(2012)\citenamefont{Pigolotti, Benzi,
  Jensen, and Nelson}}]{Pigolotti:2012tx}
\bibinfo{author}{\bibfnamefont{S.}~\bibnamefont{Pigolotti}},
  \bibinfo{author}{\bibfnamefont{R.}~\bibnamefont{Benzi}},
  \bibinfo{author}{\bibfnamefont{M.~H.} \bibnamefont{Jensen}},
  \bibnamefont{and} \bibinfo{author}{\bibfnamefont{D.~R.}
  \bibnamefont{Nelson}}, \bibinfo{journal}{Physical Review Letters}
  \textbf{\bibinfo{volume}{108}}, \bibinfo{pages}{128102}
  (\bibinfo{year}{2012}).

\bibitem[{\citenamefont{Pigolotti et~al.}(2013)\citenamefont{Pigolotti, Benzi,
  Perlekar, Jensen, Toschi, and Nelson}}]{Pigolotti:2013le}
\bibinfo{author}{\bibfnamefont{S.}~\bibnamefont{Pigolotti}},
  \bibinfo{author}{\bibfnamefont{R.}~\bibnamefont{Benzi}},
  \bibinfo{author}{\bibfnamefont{P.}~\bibnamefont{Perlekar}},
  \bibinfo{author}{\bibfnamefont{M.~H.} \bibnamefont{Jensen}},
  \bibinfo{author}{\bibfnamefont{F.}~\bibnamefont{Toschi}}, \bibnamefont{and}
  \bibinfo{author}{\bibfnamefont{D.}~\bibnamefont{Nelson}},
  \bibinfo{journal}{Theoretical Population Biology}
  \textbf{\bibinfo{volume}{84}}, \bibinfo{pages}{72} (\bibinfo{year}{2013}).

\bibitem[{\citenamefont{Korolev}(2013)}]{Korolev:2013uk}
\bibinfo{author}{\bibfnamefont{K.~S.} \bibnamefont{Korolev}},
  \bibinfo{journal}{PLoS Computational Biology} \textbf{\bibinfo{volume}{9}},
  \bibinfo{pages}{e1002994} (\bibinfo{year}{2013}).

\bibitem[{\citenamefont{Pearl and Slobodkin}(1976)}]{Pearl:1976fk}
\bibinfo{author}{\bibfnamefont{R.}~\bibnamefont{Pearl}} \bibnamefont{and}
  \bibinfo{author}{\bibfnamefont{L.}~\bibnamefont{Slobodkin}},
  \bibinfo{journal}{Quarterly Review of Biology} pp. \bibinfo{pages}{6--24}
  (\bibinfo{year}{1976}).

\bibitem[{\citenamefont{Van~Kampen}(1992)}]{Van-Kampen:1992tk}
\bibinfo{author}{\bibfnamefont{N.~G.} \bibnamefont{Van~Kampen}},
  \emph{\bibinfo{title}{Stochastic Processes in Physics and Chemistry}},
  vol.~\bibinfo{volume}{1} (\bibinfo{publisher}{Elsevier},
  \bibinfo{year}{1992}).

\bibitem[{\citenamefont{Gardiner}(1985)}]{Gardiner:1985qr}
\bibinfo{author}{\bibfnamefont{C.}~\bibnamefont{Gardiner}},
  \emph{\bibinfo{title}{Handbook of Stochastic Processes}}
  (\bibinfo{publisher}{Springer}, \bibinfo{year}{1985}).

\bibitem[{\citenamefont{Dykman et~al.}(1994)\citenamefont{Dykman, Mori, Ross,
  and Hunt}}]{Dykman:1994pz}
\bibinfo{author}{\bibfnamefont{M.}~\bibnamefont{Dykman}},
  \bibinfo{author}{\bibfnamefont{E.}~\bibnamefont{Mori}},
  \bibinfo{author}{\bibfnamefont{J.}~\bibnamefont{Ross}}, \bibnamefont{and}
  \bibinfo{author}{\bibfnamefont{P.}~\bibnamefont{Hunt}}, \bibinfo{journal}{The
  Journal of Chemical Physics} \textbf{\bibinfo{volume}{100}},
  \bibinfo{pages}{5735} (\bibinfo{year}{1994}).

\bibitem[{\citenamefont{Risken}(1984)}]{Risken:1984df}
\bibinfo{author}{\bibfnamefont{H.}~\bibnamefont{Risken}},
  \emph{\bibinfo{title}{Fokker-Planck Equation}}
  (\bibinfo{publisher}{Springer}, \bibinfo{year}{1984}).

\bibitem[{\citenamefont{Langer}(1969)}]{Langer:1969aa}
\bibinfo{author}{\bibfnamefont{J.}~\bibnamefont{Langer}},
  \bibinfo{journal}{Annals of Physics} \textbf{\bibinfo{volume}{54}},
  \bibinfo{pages}{258} (\bibinfo{year}{1969}).

\bibitem[{\citenamefont{Ovaskainen and Meerson}(2010)}]{Ovaskainen:2010bc}
\bibinfo{author}{\bibfnamefont{O.}~\bibnamefont{Ovaskainen}} \bibnamefont{and}
  \bibinfo{author}{\bibfnamefont{B.}~\bibnamefont{Meerson}},
  \bibinfo{journal}{Trends in ecology and evolution}
  \textbf{\bibinfo{volume}{25}}, \bibinfo{pages}{643} (\bibinfo{year}{2010}).

\bibitem[{\citenamefont{Crow et~al.}(1970)\citenamefont{Crow, Kimura
  et~al.}}]{Crow:1970jh}
\bibinfo{author}{\bibfnamefont{J.~F.} \bibnamefont{Crow}},
  \bibinfo{author}{\bibfnamefont{M.}~\bibnamefont{Kimura}},
  \bibnamefont{et~al.}, \emph{\bibinfo{title}{An Introduction to Population
  Genetics Theory.}} (\bibinfo{publisher}{New York, Evanston and London: Harper
  \& Row, Publishers}, \bibinfo{year}{1970}).

\bibitem[{\citenamefont{Kimura and Ohta}(1969)}]{Kimura:1969pt}
\bibinfo{author}{\bibfnamefont{M.}~\bibnamefont{Kimura}} \bibnamefont{and}
  \bibinfo{author}{\bibfnamefont{T.}~\bibnamefont{Ohta}},
  \bibinfo{journal}{Genetics} \textbf{\bibinfo{volume}{61}},
  \bibinfo{pages}{763} (\bibinfo{year}{1969}).

\bibitem[{\citenamefont{Constable and McKane}(2015)}]{Constable:2015aa}
\bibinfo{author}{\bibfnamefont{G.~W.~A.} \bibnamefont{Constable}}
  \bibnamefont{and} \bibinfo{author}{\bibfnamefont{A.~J.}
  \bibnamefont{McKane}}, \bibinfo{journal}{Physical Review Letters}
  \textbf{\bibinfo{volume}{114}}, \bibinfo{pages}{038101}
  (\bibinfo{year}{2015}).

\bibitem[{\citenamefont{Parsons and Quince}(2007)}]{Parsons:2007aa}
\bibinfo{author}{\bibfnamefont{T.~L.} \bibnamefont{Parsons}} \bibnamefont{and}
  \bibinfo{author}{\bibfnamefont{C.}~\bibnamefont{Quince}},
  \bibinfo{journal}{Theoretical Population Biology}
  \textbf{\bibinfo{volume}{72}}, \bibinfo{pages}{468} (\bibinfo{year}{2007}).

\bibitem[{\citenamefont{Parsons et~al.}(2008)\citenamefont{Parsons, Quince, and
  Plotkin}}]{Parsons:2008aa}
\bibinfo{author}{\bibfnamefont{T.~L.} \bibnamefont{Parsons}},
  \bibinfo{author}{\bibfnamefont{C.}~\bibnamefont{Quince}}, \bibnamefont{and}
  \bibinfo{author}{\bibfnamefont{J.~B.} \bibnamefont{Plotkin}},
  \bibinfo{journal}{Theoretical Population Biology}
  \textbf{\bibinfo{volume}{74}}, \bibinfo{pages}{302} (\bibinfo{year}{2008}).

\bibitem[{\citenamefont{Kogan et~al.}(2014)\citenamefont{Kogan, Khasin,
  Meerson, Schneider, and Myers}}]{Kogan:2014aa}
\bibinfo{author}{\bibfnamefont{O.}~\bibnamefont{Kogan}},
  \bibinfo{author}{\bibfnamefont{M.}~\bibnamefont{Khasin}},
  \bibinfo{author}{\bibfnamefont{B.}~\bibnamefont{Meerson}},
  \bibinfo{author}{\bibfnamefont{D.}~\bibnamefont{Schneider}},
  \bibnamefont{and} \bibinfo{author}{\bibfnamefont{C.~R.} \bibnamefont{Myers}},
  \bibinfo{journal}{Physical Review E} \textbf{\bibinfo{volume}{90}},
  \bibinfo{pages}{042149} (\bibinfo{year}{2014}).

\bibitem[{\citenamefont{Lin et~al.}(2012)\citenamefont{Lin, Kim, and
  Doering}}]{Lin:2012wf}
\bibinfo{author}{\bibfnamefont{Y.~T.} \bibnamefont{Lin}},
  \bibinfo{author}{\bibfnamefont{H.}~\bibnamefont{Kim}}, \bibnamefont{and}
  \bibinfo{author}{\bibfnamefont{C.~R.} \bibnamefont{Doering}},
  \bibinfo{journal}{Journal of Statistical Physics}
  \textbf{\bibinfo{volume}{148}}, \bibinfo{pages}{647} (\bibinfo{year}{2012}).

\bibitem[{\citenamefont{Kleinert}(2009)}]{Kleinert:2009aa}
\bibinfo{author}{\bibfnamefont{H.}~\bibnamefont{Kleinert}},
  \emph{\bibinfo{title}{Path Integrals in Quantum Mechanics, Statistics,
  Polymer Physics, and Financial Markets}} (\bibinfo{publisher}{World
  Scientific}, \bibinfo{year}{2009}).

\bibitem[{\citenamefont{Altland and Simons}(2010)}]{Altland:2010aa}
\bibinfo{author}{\bibfnamefont{A.}~\bibnamefont{Altland}} \bibnamefont{and}
  \bibinfo{author}{\bibfnamefont{B.~D.} \bibnamefont{Simons}},
  \emph{\bibinfo{title}{Condensed Matter Field Theory}}
  (\bibinfo{publisher}{Cambridge University Press}, \bibinfo{year}{2010}).

\bibitem[{\citenamefont{Gillespie}(1976)}]{Gillespie:1976km}
\bibinfo{author}{\bibfnamefont{D.~T.} \bibnamefont{Gillespie}},
  \bibinfo{journal}{Journal of Computational Physics}
  \textbf{\bibinfo{volume}{22}}, \bibinfo{pages}{403} (\bibinfo{year}{1976}).

\bibitem[{\citenamefont{Gillespie}(1977)}]{Gillespie:1977fk}
\bibinfo{author}{\bibfnamefont{D.~T.} \bibnamefont{Gillespie}},
  \bibinfo{journal}{The Journal of Physical Chemistry}
  \textbf{\bibinfo{volume}{81}}, \bibinfo{pages}{2340} (\bibinfo{year}{1977}).

\bibitem[{\citenamefont{Grasman and Herwaarden}(1999)}]{Grasman:1999kx}
\bibinfo{author}{\bibfnamefont{J.}~\bibnamefont{Grasman}} \bibnamefont{and}
  \bibinfo{author}{\bibfnamefont{O.~A.} \bibnamefont{Herwaarden}},
  \emph{\bibinfo{title}{Asymptotic Methods for the Fokker-Planck Equation and the Exit Problem in Applications}} (\bibinfo{publisher}{Springer},
  \bibinfo{year}{1999}).

\bibitem[{\citenamefont{Verhulst}(2006)}]{Verhulst:2006fk}
\bibinfo{author}{\bibfnamefont{F.}~\bibnamefont{Verhulst}},
  \emph{\bibinfo{title}{Methods and Applications of Singular Perturbations:
  Boundary Layers and Multiple Timescale Dynamics}}
  (\bibinfo{publisher}{Springer}, \bibinfo{year}{2006}).

\bibitem[{\citenamefont{Bender and Orszag}(1999)}]{Bender:1999uq}
\bibinfo{author}{\bibfnamefont{C.~M.} \bibnamefont{Bender}} \bibnamefont{and}
  \bibinfo{author}{\bibfnamefont{S.~A.} \bibnamefont{Orszag}},
  \emph{\bibinfo{title}{Advanced Mathematical Methods for Scientists and
  Engineers I: Asymptotic Methods and Perturbation Theory}},
  vol.~\bibinfo{volume}{1} (\bibinfo{publisher}{Springer},
  \bibinfo{year}{1999}).

\bibitem[{\citenamefont{Hinch}(1991)}]{Hinch:xy}
\bibinfo{author}{\bibfnamefont{E.}~\bibnamefont{Hinch}},
  \emph{\bibinfo{title}{Perturbation Methods. 1991}}.

\bibitem[{\citenamefont{Assaf and Meerson}(2010)}]{Assaf:2010xp}
\bibinfo{author}{\bibfnamefont{M.}~\bibnamefont{Assaf}} \bibnamefont{and}
  \bibinfo{author}{\bibfnamefont{B.}~\bibnamefont{Meerson}},
  \bibinfo{journal}{Physical Review E} \textbf{\bibinfo{volume}{81}},
  \bibinfo{pages}{021116} (\bibinfo{year}{2010}).

\bibitem[{\citenamefont{Kessler and Shnerb}(2007)}]{Kessler:2007bj}
\bibinfo{author}{\bibfnamefont{D.~A.} \bibnamefont{Kessler}} \bibnamefont{and}
  \bibinfo{author}{\bibfnamefont{N.~M.} \bibnamefont{Shnerb}},
  \bibinfo{journal}{Journal of Statistical Physics}
  \textbf{\bibinfo{volume}{127}}, \bibinfo{pages}{861} (\bibinfo{year}{2007}).

\bibitem[{\citenamefont{Elgart and Kamenev}(2004)}]{Elgart:2004ya}
\bibinfo{author}{\bibfnamefont{V.}~\bibnamefont{Elgart}} \bibnamefont{and}
  \bibinfo{author}{\bibfnamefont{A.}~\bibnamefont{Kamenev}},
  \bibinfo{journal}{Physical Review E} \textbf{\bibinfo{volume}{70}},
  \bibinfo{pages}{041106} (\bibinfo{year}{2004}).

\bibitem[{\citenamefont{Kamenev and Meerson}(2008)}]{Kamenev:2008xu}
\bibinfo{author}{\bibfnamefont{A.}~\bibnamefont{Kamenev}} \bibnamefont{and}
  \bibinfo{author}{\bibfnamefont{B.}~\bibnamefont{Meerson}},
  \bibinfo{journal}{Physical Review E} \textbf{\bibinfo{volume}{77}},
  \bibinfo{pages}{061107} (\bibinfo{year}{2008}).

\bibitem[{\citenamefont{Melbinger et~al.}(2010)\citenamefont{Melbinger, Cremer,
  and Frey}}]{Melbinger:2010wq}
\bibinfo{author}{\bibfnamefont{A.}~\bibnamefont{Melbinger}},
  \bibinfo{author}{\bibfnamefont{J.}~\bibnamefont{Cremer}}, \bibnamefont{and}
  \bibinfo{author}{\bibfnamefont{E.}~\bibnamefont{Frey}},
  \bibinfo{journal}{Physical Review Letters} \textbf{\bibinfo{volume}{105}},
  \bibinfo{pages}{178101} (\bibinfo{year}{2010}).

\bibitem[{\citenamefont{Cremer et~al.}(2011)\citenamefont{Cremer, Melbinger,
  and Frey}}]{Cremer:2011pi}
\bibinfo{author}{\bibfnamefont{J.}~\bibnamefont{Cremer}},
  \bibinfo{author}{\bibfnamefont{A.}~\bibnamefont{Melbinger}},
  \bibnamefont{and} \bibinfo{author}{\bibfnamefont{E.}~\bibnamefont{Frey}},
  \bibinfo{journal}{Physical Review E} \textbf{\bibinfo{volume}{84}},
  \bibinfo{pages}{051921} (\bibinfo{year}{2011}).

\bibitem[{\citenamefont{Van~Dyken et~al.}(2013)\citenamefont{Van~Dyken,
  M{\"u}ller, Mack, and Michael}}]{Dyken:2013ml}
\bibinfo{author}{\bibfnamefont{J.~D.} \bibnamefont{Van~Dyken}},
  \bibinfo{author}{\bibfnamefont{M.~J.} \bibnamefont{M{\"u}ller}},
  \bibinfo{author}{\bibfnamefont{K.~M.} \bibnamefont{Mack}}, \bibnamefont{and}
  \bibinfo{author}{\bibfnamefont{D.~M.} \bibnamefont{Michael}},
  \bibinfo{journal}{Current Biology} \textbf{\bibinfo{volume}{23}},
  \bibinfo{pages}{919} (\bibinfo{year}{2013}).

\bibitem[{\citenamefont{Graham and T{\'e}l}(1985)}]{Graham:1985rf}
\bibinfo{author}{\bibfnamefont{R.}~\bibnamefont{Graham}} \bibnamefont{and}
  \bibinfo{author}{\bibfnamefont{T.}~\bibnamefont{T{\'e}l}},
  \bibinfo{journal}{Physical Review A} \textbf{\bibinfo{volume}{31}},
  \bibinfo{pages}{1109} (\bibinfo{year}{1985}).

\end{thebibliography}

\end{document}